\documentclass[superscriptaddress,prb, 10pt,aps, onecolumn,  notitlepage, showpacs, floatfix]{revtex4-1}

\pdfoutput=1
\usepackage{graphicx}
\usepackage{amsmath, amssymb, amsfonts, bm}
\usepackage{latexsym}
\usepackage[colorlinks]{hyperref}
\usepackage{mathdots}

\usepackage[protrusion=true,expansion=true]{microtype}

\newcommand{\fixme}[1]%
   {\begingroup{\color{blue}\it (FIXME: #1)}\endgroup}
\newcommand{\highlight}[1]%
   {\begingroup{\color{blue} #1}\endgroup}

\begin{document}

\title{Efficient protocol for qubit initialization with a tunable environment}
\date{\today}

\author{Jani Tuorila}
\affiliation{QCD Labs, COMP Centre of Excellence, Department of Applied Physics, Aalto University, P.O. Box 13500, FI-00076 Aalto, Finland}
\affiliation{MSP group, COMP Centre of Excellence, Department of Applied Physics, Aalto University, P.O. Box 13500, FI-00076 Aalto, Finland}
\affiliation{Theoretical Physics Research Unit, University of Oulu, P.O. Box 3000, FI-90014 Oulu, Finland}
\author{Matti Partanen}
\affiliation{QCD Labs, COMP Centre of Excellence, Department of Applied Physics, Aalto University, P.O. Box 13500, FI-00076 Aalto, Finland}
\author{Tapio Ala-Nissila}
\affiliation{MSP group, COMP Centre of Excellence, Department of Applied Physics, Aalto University, P.O. Box 13500, FI-00076 Aalto, Finland}
\affiliation{Department of Physics, Brown University, Box 1843, Providence, Rhode Island 02912-1843, U.S.A.}
\author{Mikko M\"ott\"onen}
\affiliation{QCD Labs, COMP Centre of Excellence, Department of Applied Physics, Aalto University, P.O. Box 13500, FI-00076 Aalto, Finland}



\begin{abstract}
We propose an efficient qubit initialization protocol based on a dissipative environment that can be dynamically adjusted. Here the qubit is coupled to a thermal bath through a tunable harmonic oscillator. On-demand initialization is achieved by sweeping the oscillator rapidly into resonance with the qubit. This resonant coupling with the engineered environment induces fast relaxation to the ground state of the system, and a consecutive rapid sweep back to off resonance guarantees weak excess dissipation during quantum computations. We solve the corresponding quantum dynamics using a Markovian master equation for the reduced density operator of the qubit-bath system. This allows us to optimize the parameters and the initialization protocol for the qubit. Our analytical calculations show that the ground-state occupation of our system is well protected during the fast sweeps of the environmental coupling and, consequently, we obtain an estimate for the duration of our protocol by solving the transition rates between the low-energy eigenstates with the Jacobian diagonalization method. Our results suggest that the current experimental state of the art for the initialization speed of superconducting qubits at a given fidelity can be considerably improved. 
\end{abstract}

\flushbottom
\maketitle

%
%
\thispagestyle{empty}


\section*{Introduction}


Preparation of a qubit into a well-defined initial state is one of the key requirements for any quantum computational algorithm~\cite{DiVincenzo00,NielsenChuang}. The conventional passive initialization protocol relies on the relaxation of the qubit to a thermal state determined by the residual coupling to the environment. This protocol is inherently slow because the relaxation rate has to be minimized to decrease the probability of errors in a coherent quantum computation. 

In the context of error-free quantum computing, a long initialization time 
would not present a problem since the quantum register has to be prepared only once in the beginning of the computation. 
However, realistic quantum computational devices also suffer from gate errors which have to be corrected with quantum-error-correction codes~\cite{Shor95}. Such codes rely on logical qubits which consist of several physical qubits. The codes initiate from a predetermined state for the physical qubits and are being constantly executed during a computation. They also have strict requirements for the initialization and gate error thresholds for individual qubits, of the order of $10^{-5}$ for the conventional concatenated codes~\cite{Preskill98,Schindler11}. In the more refined topological quantum error correction codes~\cite{Kitaev03}, the logical error can be suppressed with stabilizing measurements which increase the thresholds up to $10^{-2}$ for the physical qubit operations and lead to improved protection of quantum information during the computation. Nevertheless, stabilizer codes such as surface~\cite{Fowler12,Kelly15} and color~\cite{Bombin06,Nigg14} codes still require a continuous supply of measurement qubits in a known low-entropy state. Thus initialization time is also an issue in large-scale quantum computing.

Fast and accurate qubit initialization remains a technological challenge in the superconducting qubit implementations which have shown great potential for scalability~\citep{Kelly15}. Major developments have been made with active protocols, such as initialization by successive projective measurements~\cite{Riste12, Johnson12,Govia15}, by Purcell-filtered cavity~\cite{Reed10}, or by cooling with a coherent microwave drive~\cite{Valenzuela06,Grajcar08,Geerlings13, Jin15}. 
The typical figure of merit of a protocol is the time $\tau_{10}$ required for a qubit excitation to decay by an order of magnitude. In the reported experiments~\cite{Reed10}, a thermal-equilibrium fidelity of $99.9\%$ with $\tau_{10}= 40$ ns has been obtained with a protocol for a Purcell-filtered superconducting qubit with transition frequency $\omega_0/(2\pi)=5.16$ GHz and intrinsic relaxation time $T_1 = 540$ ns. However, the method uses a tuned qubit which is not desirable since it reserves a broad frequency band, and a high ground-state fidelity using a long-lived qubit remains to be demonstrated with this method. An initialization protocol based on coherent driving has reached a $99.5 \%$ ground-state fidelity with $\tau_{10}=1.4 \ \mu$s, $\omega_0/(2\pi)= 5$ GHz, and $T_1=37 \ \mu$s~\cite{Geerlings13}, without relying on qubit tuning. Even though this protocol meets the error thresholds of the topological codes, large-scale quantum computing calls for shorter initialization times. 
Furthermore, initialization fidelities greater than 99.99$\%$ are preferred in large-scale computations to reduce the number of physical qubits needed for a logical qubit. 

In solid-state systems, an initialization protocol can potentially be realized by strongly coupling the qubit to a cold dissipative reservoir for fast initialization and by switching off the coupling for the actual computation\cite{StolzeSuter}. If the bath has a lower effective temperature than the qubit, the ground state fidelity of the qubit is increased. This type of setup is a part of environmental quantum-state engineering by dissipation, where one aims to drive the system into a desired steady state by using a carefully tailored environment\cite{Poyatos96,Verstraete09,Clark03,Kastoryano11,Rao14}. Such engineering has already been used in generation of coherent superposition states~\cite{Murch12,Kienzler15,Leghtas15}, in creation of entanglement\cite{Krauter11,Lin13,Shankar13}, and in simulations of open quantum systems~\cite{Barreiro11,Toth16}. 

\begin{figure}[t]
\centering
\includegraphics[width=0.50\linewidth]{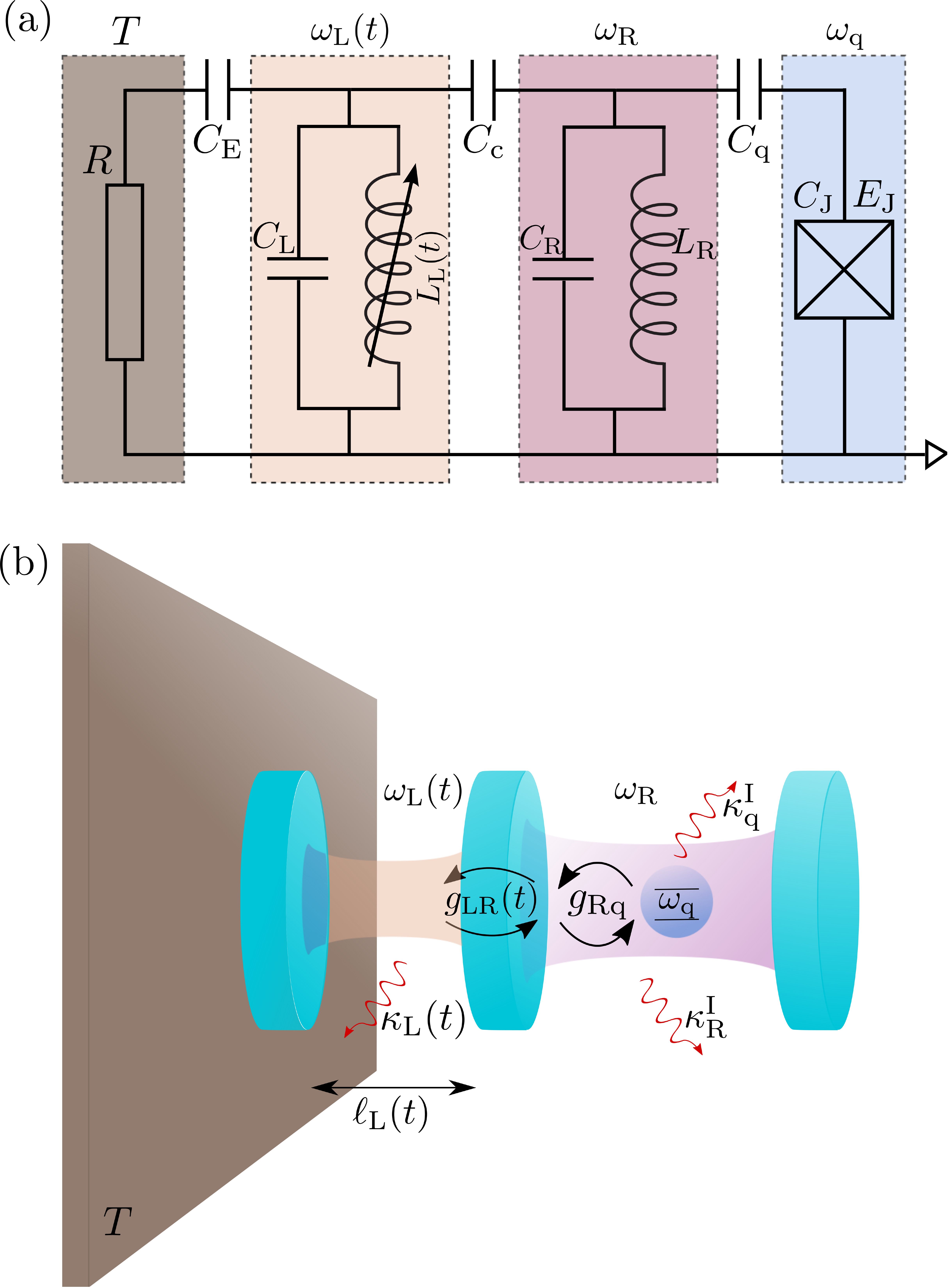}
\caption{Schematic qubit initialization setup. (a) Lumped-element circuit model. A superconducting qubit (blue box) with angular frequency $\omega_{\rm q}$ and intrinsic relaxation rate $\kappa_{\rm q}^{\rm I}$ is indirectly coupled to a thermal bath (brown), formed by a resistor $R$, through two $LC$ resonators. By tuning the inductance $L_{\rm L}(t)$ of the left resonator (orange), one can tune its bare resonance frequency and coupling strengths $g_{\rm LR}(t)$ and $\kappa_{\rm L}(t)$ with the right resonator (magenta) and the bath (temperature $T$), respectively. The right resonator has a bare angular frequency of $\omega_{\rm R}$ and is coupled to the qubit and an intrinsic bath with coupling strengths $g_{\rm Rq}$ and $\kappa_{\rm R}^{\rm I}$. (b) Analogous cavity QED setup where a two-level atom is coupled to a thermal bath through two optical cavities. The coupling to the thermal bath is controlled by tuning the length $\ell_{\rm L}(t)$ of the left cavity.}
\label{fig:schema}
\end{figure}

In this paper, we focus on a ground-state initialization proposal~\cite{Jones13a,Jones13b}, where a superconducting qubit and a low-temperature resistive bath are coupled indirectly through two resonators as shown in Figure~\ref{fig:schema}. The effective inductance of the left resonator, which is capacitively coupled to the bath, can be dynamically adjusted, allowing control over its bare resonance frequency. If the left resonator is sufficiently detuned from the qubit, it shunts the noise of the resistive bath at the qubit frequency, and hence the decoherence of the qubit is dictated by its slow intrinsic relaxation rate. If the left resonator is in resonance with the qubit, the qubit couples strongly to the bath leading to an increased relaxation rate. Previous calculations of static transition rates in this scenario indicate that the lifetime of the qubit can be controlled over several orders of magnitude~\cite{Jones13a,Jones13b}. However, an actual initialization protocol and its dynamics, speed, and fidelity have not been reported. 
Here, we develop a qubit initialization protocol and solve its dynamics by using a Markovian master equation. We show analytically that the ground state is protected during the sweeps of the left resonator to and from the resonance, implying that the speed of the protocol is determined by the strong resonant coupling with the dissipative bath. At the resonance, we find an approximative analytic solution for the low-energy eigenproblem using Jacobian diagonalization~\cite{Jacobi1846, NR11}. It yields a useful lower bound for the duration of the protocol at a given fidelity. Optimization of the physical circuit parameters 
suggests that, with present-day technology, the current benchmarks for fidelity and protocol speed~\cite{Geerlings13} can be considerably improved with the help of our protocol.


\section*{Results}


\subsection*{System}

The above-discussed tunable-environment qubit can be conveniently studied using a lumped-element circuit model shown in Figure~\ref{fig:schema}(a). Here, the superconducting qubit with the transition energy $\hbar\omega_\textrm{q}$ is coupled to a bosonic heat bath through two $LC$ resonators. Both resonators are formed by a  lumped capacitance $C_k$ and an inductance $L_k$, where $k=\textrm{L,R}$ refer to the left and right resonators, respectively. The bath arises from the resistance $R$ at temperature $T$. The left resonator is coupled directly to the bath and to the right resonator through capacitances $C_\textrm{E}$ and $C_\textrm{c}$, respectively. The inductance $L_\textrm{L}(t)$ of the left $LC$ resonator is made tunable using a SQUID, the Josephson inductance of which is controlled by an external magnetic flux. As a consequence, the bare angular frequency of the left resonator  $\omega_\textrm{L}(t)=1/\sqrt{L_\textrm{L}(t)C_\textrm{L}}$ can also be adjusted with the external flux. The right resonator has a fixed bare angular frequency $\omega_\textrm{R}=1/\sqrt{L_\textrm{R}C_\textrm{R}}$ and is coupled to the qubit through the capacitance $C_\textrm{q}$. 

The Hamiltonian of the circuit can be written as~\cite{Jones13b}
\begin{equation}\label{eq:totalHam}
\hat{H}(t) = \hat{H}_{\rm S}(t) + \hat{H}_{\rm E} + \hat{H}_{\rm I}(t) ,
\end{equation}
where the three terms describe the system, the resistive environment, and their interaction, respectively. The system Hamiltonian can be expressed as
\begin{equation}
\hat{H}_{\rm S}(t) = \hat{H}_0(t) + \hat{H}_1(t),\label{eq:HS}
\end{equation}
where
\begin{equation}\label{eq:uncoupH}
\hat{H}_0(t) = \hbar \omega_{\rm L}(t) \hat{a}_{\rm L}^{\dag}\hat{a}_{\rm L} + \hbar \omega_{\rm R} \hat{a}_{\rm R}^{\dag}\hat{a}_{\rm R} + \hbar \omega_{\rm q} \hat{\sigma}_+\hat{\sigma}_-,
\end{equation}
and
\begin{eqnarray}
\hat{H}_1(t) &=& \hbar g_{\rm LR}(t)(\hat{a}_{\rm L}^{\dag}+\hat{a}_{\rm L})(\hat{a}_{\rm R}^{\dag}+\hat{a}_{\rm R}) + i\hbar g_{\rm Rq}(\hat{a}_{\rm R}^{\dag}+\hat{a}_{\rm R})(\hat{\sigma}_- - \hat{\sigma}_+)\\
&\approx & \hbar g_{\rm LR}(t)(\hat{a}_{\rm L}^{\dag}\hat{a}_{\rm R}+\hat{a}_{\rm L}\hat{a}_{\rm R}^{\dag}) + i\hbar g_{\rm Rq}(\hat{a}_{\rm R}^{\dag}\hat{\sigma}_- - \hat{a}_{\rm R}\hat{\sigma}_+)\label{eq:RWAHam}.
\end{eqnarray}
Above, $\hat{a}_\textrm{L}$, $\hat{a}_\textrm{R}$, and $\hat{\sigma}_-$ are the annihilation operators for the left and right resonators and the qubit, respectively. In the following, we denote the eigenstates of the unperturbed Hamiltonian~(\ref{eq:uncoupH}) with $|n_{\rm L},n_{\rm R},n_{\rm q}\rangle$, where the occupation numbers of the left and right resonators can have values $n_{\rm L},n_{\rm R}=0,1,2,\ldots$ and that of the qubit assumes values $n_{\rm q} = \textrm{g,e}$, which stand for ground and excited state, respectively. The Hamiltonian $\hat{H}_{\rm S}(t)$ describes a tripartite system formed by two harmonic resonators and a qubit. The right resonator is coupled bi-linearly to the left resonator and to the qubit with the respective coupling frequencies $g_{\rm LR}(t) = g_{\rm LR}^0\sqrt{\omega_\textrm{L}(t)/\omega_{\rm R}}$ and $g_{\rm Rq} = \frac{e C_{\rm q}\sqrt{\hbar \omega_{\rm R}/C_{\rm R}}}{\hbar(C_{\rm q}+C_{\rm J})}\left(\frac{E_{\rm J}}{E_{\rm C}}\right)^{1/4}$, where $g_{\rm LR}^0 = \omega_{\rm R}C_{\rm c}/(2\sqrt{C_{\rm L}C_{\rm R}})$ is the resonant coupling strength between the left and right resonators, $C_{\rm J}$ is the capacitance of the Josephson junction, and $E_{\rm J}$ and $E_{\rm C}=e^2/[2(C_{\rm q}+C_{\rm J})]$ are the Josephson coupling energy and the charging energy per electron for the superconducting island, respectively. 
Coupling between the qubit and the left resonator is mediated by the right resonator and is, therefore, of second order in coupling frequencies $g_{\rm LR}^0$ and $g_{\rm Rq}$. Consequently, the right resonator acts as an additional filter for the thermal noise of the bath. In our analytic considerations, we apply the rotating-wave approximation (RWA) for both of the coupling terms, cf. Equation~(\ref{eq:RWAHam}).
The interaction with the bath is also bi-linear and described by
\begin{equation}\label{eq:intHam}
\hat{H}_{\rm I}(t) = -C_{\rm E}\hat{V}_{\rm L}\delta \hat{V}_{\rm res}=-C_{\rm E}\sqrt{\frac{\hbar\omega_{\rm L}(t)}{2C_{\rm L}}}(\hat{a}_{\rm L}^{\dag}+\hat{a}_{\rm L})\delta \hat{V}_{\rm res},
\end{equation}
where $\hat{V}_{\rm L}=\sqrt{\frac{\hbar\omega_{\rm L}(t)}{2C_{\rm L}}}(\hat{a}_{\rm L}^{\dag}+\hat{a}_{\rm L})$ is the voltage over the left resonator, and $\delta \hat{V}_{\rm res}$ describes the voltage fluctuations over the resistor. The resistor Hamiltonian $\hat{H}_{\rm E}$ is given by that of a bosonic thermal bath~\cite{Caldeira81}. We do not express it explicitly since we are only interested in the transition rates which are determined in thermal equilibrium by the Johnson--Nyquist spectrum of the voltage fluctuations:
\begin{equation}
S_{\delta V_{\rm res}}(\omega)= \frac{2\hbar R \omega}{1-e^{-\hbar\omega/(k_{\rm B}T)}}.
\end{equation}
Furthermore, we neglect any filtering of this spectrum owing to the resistor itself by assuming that the frequencies relevant for the system obey $\omega\ll 1/(RC_{\rm R})$~\cite{Jones13b}.

We emphasize that the following discussion is not specific to the lumped-element model or superconducting qubits, but can be used rather generally for qubits with indirect adjustable coupling to a thermal bath, as shown in Ref.~\onlinecite{Jones13b} with the distributed circuit elements which are frequently used in contemporary circuit QED.

\subsection*{Transition rates}

The transition rates from the $m$th instantaneous eigenstate of the Hamilonian $\hat{H}_{\rm S}(t)$ to the $n$th state can be calculated from Fermi's golden rule as
\begin{eqnarray}
\Gamma_{mn}(t) &=& \frac{|\langle \Psi_n(t)|C_{\rm E}\hat{V}_{\rm L}|\Psi_m(t)\rangle|^2}{\hbar^2}S_{\delta V_{\rm res}}\left[-\omega_{mn}(t)\right]\nonumber\\
&=& \Gamma_0 |\langle \Psi_n(t)|(\hat{a}_{\rm L}^{\dag}+\hat{a}_{\rm L})|\Psi_m(t)\rangle|^2\frac{\omega_{\rm L}\omega_{nm}(t)}{\omega_{\rm R}^2}\frac{1}{1-e^{-\hbar\omega_{nm}(t)/(k_{\rm B}T)}},\label{eq:statrate}
\end{eqnarray}
where $\omega_{mn}(t)=\omega_n(t)-\omega_m(t)$, and $\omega_k(t)$ are the eigenfrequencies corresponding to the eigenstates $|\Psi_k(t)\rangle$ of the Hamiltonian $\hat{H}_{\rm S}(t)$, i.e. $\hat{H}_{\rm S}(t)|\Psi_k(t)\rangle = \hbar \omega_k(t)|\Psi_k(t)\rangle$, $\Gamma_0 = (C_{\rm E}/\sqrt{C_{\rm L}C_{\rm R}})^2(R/Z_{\rm R})\omega_{\rm R}$, and $Z_{\rm R}=\sqrt{L_{\rm R}/C_{\rm R}}$. We thus observe that positive-frequency fluctuations in the environment cause emission in the small system~\cite{Clerk10}.
Note, that in Equation~(\ref{eq:statrate}) we express the transition rates in units of $\Gamma_0$, which equals the bare zero-temperature left-resonator transition rate for $\omega_{\rm L}(t)=\omega_{\rm R}$.  Clearly, we can maximize the transition rates by maximizing $C_{\rm E}$ and $R$ with respect to $C_{\rm R}$ and $Z_{\rm R}$, respectively. We further note that in principle $C_{\rm E}$ should be added to $C_{\rm L}$ to obtain the effective left-resonator capacitance, but if $C_{\rm E}\ll C_{\rm L}$ its effects on the eigenstates and eigenenergies of the system are negligible~\cite{Jones13b}.

\begin{figure}[ht]
\centering
\includegraphics[width=0.8\linewidth]{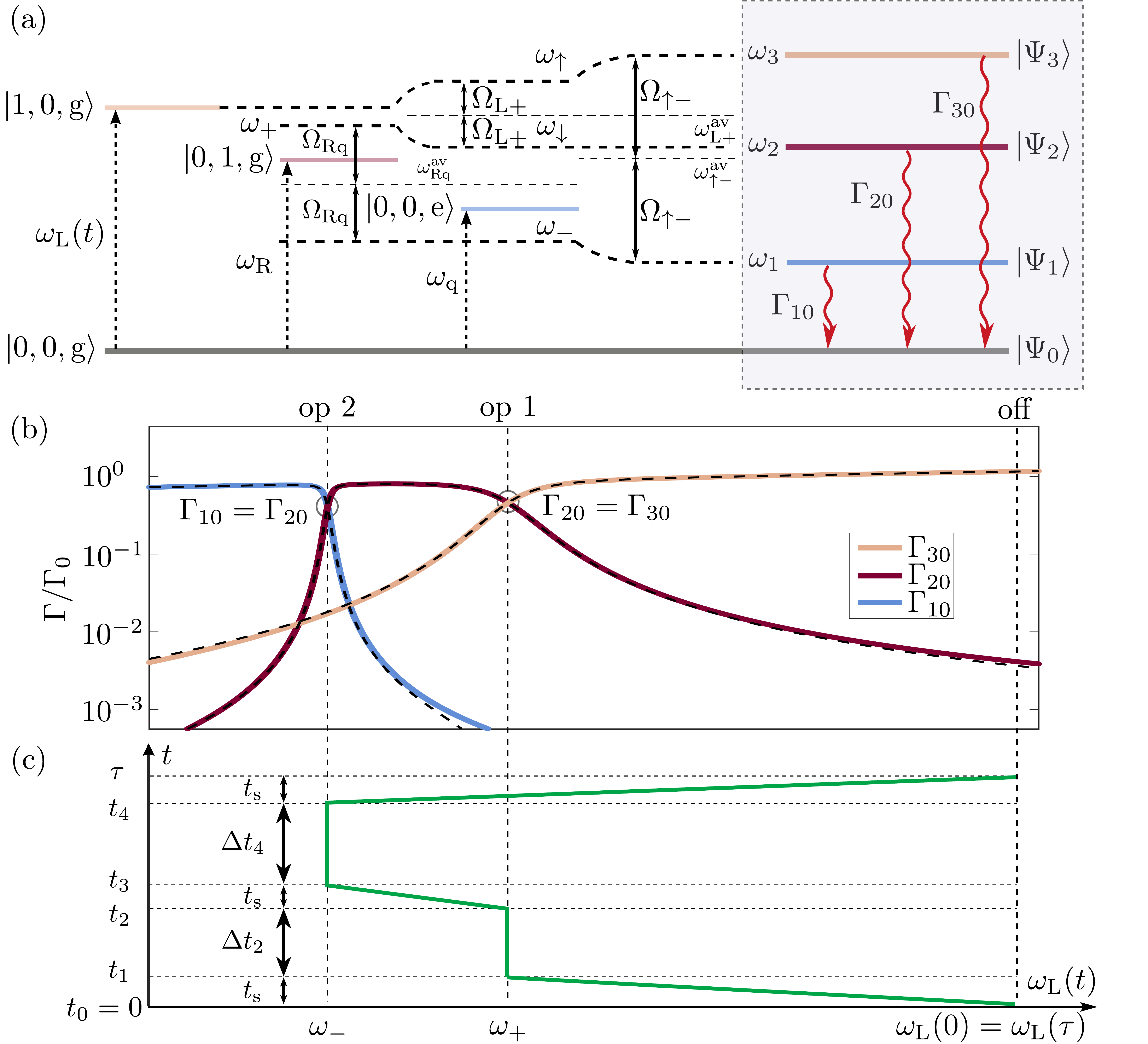}
\caption{Low-energy level structure, relaxation rates, and the proposed initialization protocol. (a) Energy levels in the Jacobian diagonalization scheme for the four lowest eigenstates. We make three Jacobian diagonalizing transformations in the subspaces $\{|0,0,\textrm{e}\rangle , |0,1,\textrm{g}\rangle \}$, $\{|1,0,\textrm{g}\rangle , |+\rangle \}$, and $\{|-\rangle , |\uparrow \rangle \}$ resulting in the approximative eigenenergies $\omega_1 = \omega_{\uparrow -}^{\rm av}-\Omega_{\uparrow -}$, $\omega_2 = \omega_{\downarrow}$, and $\omega_3 = \omega_{\uparrow -}^{\rm av}+\Omega_{\uparrow -}$. The angular frequencies $\omega_{\pm}$, $\omega_{\uparrow\!/\!\downarrow}$, $\omega_{ij}^{\rm av}$, and $\Omega_{ij}$ are defined in \hyperref[sec:Methods]{Methods}. (b) Relaxation rates of the lowest three excited states as functions of the bare left-resonator angular frequency and the initialization protocol (black arrows). We plot with dashed lines the analytic rates in Equations~(\ref{eq:transrate10})--(\ref{eq:transrate30}) resulting from the Jacobian diagonalization on top of the those resulting from numerical diagonalization (solid). In the protocol, 'op' stands for operation point and 'off' refers to the state where the coupling between the qubit and the engineered environment is essentially absent. We use the parameters $C_{\rm L}=C_{\rm R} = 1$ pF, $L_{\rm R} = 250$ pH, $C_{\rm c}=C_{\rm q}=15$ fF, $E_{\rm J}/E_{\rm C} = 50$, $C_{\rm J}=26$~fF, $R= 500 \ \Omega$, $T=10$ mK, and $C_{\rm E}=4$~fF. These imply $\omega_{\rm R}/(2\pi) = 10$ GHz, $\omega_{\rm q}/(2\pi) = 9.5$ GHz, $g_{\rm LR}^0/(2\pi) = 74$ MHz, $g_{\rm Rq}/(2\pi) = 68$ MHz, and $\Gamma_0= 31\times 10^6$ s$^{-1}$. (c) Initialization protocol in terms of the control parameter $\omega_{\rm L}(t)$ (see details from the text).}
\label{fig:stream}
\end{figure}

In \hyperref[sec:Methods]{Methods}, we analytically solve the transition rates for the first three excited eigenstates employing the RWA and the Jacobian diagonalization~\citep{Jacobi1846,NR11}. After three Jacobian transformations, we obtain
\begin{eqnarray}
\Gamma_{10}^{\!\uparrow\!/\!\downarrow}(t) &=& \Gamma_0 \frac{\left[\Omega_{\rm L+}(t)\pm\delta_{\rm L+}(t)/2\right]\left[\Omega_{\!\uparrow\!/\!\downarrow -}(t)-\delta_{\!\uparrow\!/\!\downarrow -}(t)/2\right]}{4\Omega_{\rm L+}(t)\Omega_{\!\uparrow\!/\!\downarrow -}(t)}\left[\frac{\omega_{\rm L}(t)\omega_{1}^{\!\uparrow\!/\!\downarrow}(t)}{\omega_{\rm R}^2}\right] \frac{1}{1-e^{-\hbar\omega_{1}^{\!\uparrow\!/\!\downarrow}(t)/(k_{\rm B} T)}};\label{eq:transrate10}\\
\Gamma_{20}(t) &=& \Gamma_0 \frac{\left[\Omega_{\rm L+}(t)-\delta_{\rm L+}(t)/2\right]\left[\Omega_{\downarrow -}(t)+\delta_{\downarrow -}(t)/2\right]}{4\Omega_{\rm L+}(t)\Omega_{\downarrow -}(t)}\left[\frac{\omega_{\rm L}(t)\omega_2(t)}{\omega_{\rm R}^2}\right] \frac{1}{1-e^{-\hbar\omega_2(t)/(k_{\rm B} T)}}; \\
\Gamma_{30}(t) &=&\Gamma_0 \frac{[\Omega_{\rm L+}(t)+\delta_{\rm L+}(t)/2][\Omega_{\uparrow -}(t)+\delta_{\uparrow -}(t)/2]}{4\Omega_{\rm L+}(t)\Omega_{\uparrow -}(t)}\left[\frac{\omega_{\rm L}(t)\omega_{3}(t)}{\omega_{\rm R}^2}\right] \frac{1}{1-e^{-\hbar\omega_{3}(t)/(k_{\rm B} T)}},\label{eq:transrate30}
\end{eqnarray}
where $\omega_1^{\!\uparrow\!/\!\downarrow}(t)=\omega_{\!\uparrow\!/\!\downarrow -}^{\rm av}(t)-\Omega_{\!\uparrow\!/\!\downarrow -}(t)$, $\omega_2(t)=\omega_{\downarrow -}^{\rm av}(t)+\Omega_{\downarrow -}(t)$, and $\omega_3(t) = \omega_{\uparrow -}^{\rm av}(t)+ \Omega_{\uparrow -}(t)$ (see Figure~\ref{fig:stream}). We have also used the shorthand notations $\omega_{ij}^{\rm av}(t)=[\omega_{i}(t)+\omega_j(t)]/2$, $\Omega_{ij}(t) = \sqrt{[\delta_{ij}(t)/2]^2+G_{ij}(t)^2}$, $\delta_{ij}(t)=\omega_{i}(t)-\omega_j(t)$, $\omega_{\pm}=\omega_{\rm Rq}^{\rm av}\pm \Omega_{\rm Rq}$, $G_{\rm Rq} = g_{\rm Rq}$, $G_{\rm L \pm}(t)= g_{\rm LR}(t)\sqrt{[1\pm\delta_{\rm Rq}/(2\Omega_{\rm Rq})]/2}$, and $G_{\downarrow -}(t)=G_{\rm L-}(t)\sqrt{[1-\delta_{\rm L+}(t)/(2\Omega_{\rm L+}(t)]/2}$. The analytic transition rate $\Gamma_{10}(t)$ is considered separately in the two regions $\omega_{\rm L}(t)<\omega_+$ ($\downarrow$) and $\omega_{\rm L}(t)>\omega_+$ ($\uparrow$) in order the take the coupling between the qubit and the left resonator correctly into account (see \hyperref[sec:Methods]{Methods}). Note that the transition rates between the first three excited states are zero in the RWA due to the selection rules for our environmental coupling term. Furthermore, the principle of detailed balance $\Gamma_{mn} = \exp[-\hbar\omega_{mn}/(k_{\rm B} T)]\Gamma_{nm}$ holds, which implies that the excitation rates are strongly suppressed in the low temperature limit. In addition to the engineered environment described by the resistor $R$, the qubit and the right resonator typically dissipate energy to their own intrinsic environments with the transition rates $\kappa_{\rm q}^{\rm I}$ and $\kappa_{\rm R}^{\rm I}$, respectively. 

We compare the analytic rates with the corresponding numerical results in Figure~\ref{fig:stream}(b). For large detuning ($g_{\rm LR}^0,g_{\rm Rq}\ll |\omega_{\rm R}-\omega_{\rm q}|$), we find that already the third diagonalization step in the Jacobian diagonalization procedure results in a very good agreement with the numerically obtained transition rates. We also observe that when the left resonator is in resonance with either the right resonator or the qubit [$\omega_{\rm L}(t)\approx\omega_{\pm}$], the transition rates from the resonant eigenstates to the ground state are equal. We take advantage of this fact in our protocol below. In the vicinity of the resonances, there exist regions where the transition rates from the nearly resonant states change by several orders of magnitude. The widths of these regions are directly proportional to the coupling term $g_{\rm LR}^0$, and also to $g_{\rm Rq}$ for the case of $\omega_{\rm L}(t)=\omega_-$. 

\subsection*{Protocol}

Our proposed initialization protocol is depicted in Figure~\ref{fig:stream}(c) and proceeds as follows. In the beginning of the protocol ($t=0$), the parameters of the setup follow the hierarchy $g_{\rm Rq}<g_{\rm LR}^0\ll\omega_{\rm q}<\omega_{\rm R}\ll \omega_{\rm L}(0)$. In particular, the qubit frequency is chosen to be the smallest in order to minimize the effects of the possible multi-photon processes in the qubit caused by the right-resonator at any stage of the protocol. For $t=0$ and $t=\tau$, where $\tau$ is the duration of the protocol, the coupling to the engineered environment should be minimal so that the intrinsic sources of dissipation are dominating the qubit decoherence. Therefore, we refer to the bare left-resonator frequency $\omega_{\rm L}(0)=\omega_{\rm L}(\tau)$ as the 'off' state of our setup, and choose its value such that $\Gamma_{10}(0) \ll \kappa_{\rm q}^{\rm I}$ and $\Gamma_{20}(0)\ll \kappa_{\rm R}^{\rm I}$ 
(see Figures~\ref{fig:schema} and~\ref{fig:stream}). Additionally, since the widths of the regions for enhanced transition rates are proportional to $g_{\rm LR}^0$, we require that in the off state the system is in the dispersive regime where $g_{\rm LR}(0) \ll \omega_{\rm L}(0)-\omega_{\rm R}$. We also choose $g_{\rm Rq}\ll \omega_{\rm R}-\omega_{\rm q}$, which implies only weak hybridization between the qubit and the right-resonator. 

In the first stage of the protocol, the left resonator is swept fast to resonance with the effective right resonator at $\omega_{\rm L}(t_1)=\omega_+\approx \omega_{\rm R} + g_{\rm Rq}^2/\delta_{\rm Rq}$ which is only slightly hybridized with the qubit due to the dispersive coupling. As a consequence, the right resonator becomes strongly coupled to the cold bath resulting in an increase in its relaxation rate to the ground state by orders of magnitude (see Figure~\ref{fig:stream}). This operation point (denoted in Figure~\ref{fig:stream} with 'op 1') guarantees equal relaxation rates for both resonators, which is important as the relative occupations are typically not known in the beginning of the protocol and, also, because non-adiabatic transfer of occupation between the resonators can occur during the fast sweep. Any occupation in either resonator is then dissipated to the resistor during $t:t_1\rightarrow t_2$. The wait time $\Delta t_2$ depends on the required protocol error $\alpha=1-P_0(\tau)$, where we use the notation $\Delta t_i = t_i-t_{i-1}$ for the relevant protocol time intervals with $t_0=0$ and $t_5=\tau$, and define $P_0(\tau)$ as the desired ground-state occupation and the end of the protocol.

After the first relaxation step ($t\geq t_2$), the resonators are in the ground state within the protocol error $\alpha$. 
Then, the left resonator is swept into resonance with the effective qubit at $\omega_{\rm L}(t_3)=\omega_-\approx \omega_{\rm q} - g_{\rm Rq}^2/\delta_{\rm Rq}$. At this operation point (denoted in Figure~\ref{fig:stream} with 'op 2'), the relaxation rate of the qubit is increased by several orders of magnitude and is equal to the left-resonator rate. The latter is important since at this operation point, the qubit is also hybridized with the left resonator and, consequently, any initial occupation in the qubit is partly transferred to the left resonator during the sweep. By waiting for $t:t_3\rightarrow t_4$, the hybridized qubit and left resonator dissipate their energy to the resistor. The second wait time $\Delta t_4$ is also determined by the desired value of $\alpha$. Finally, the left-resonator frequency is swept back to its high initial value $\omega_{\rm L}(\tau)=\omega_{\rm L}(0)$. In addition to the wait times $\Delta t_2$ and $\Delta t_4$, the protocol duration $\tau = \sum_{i=1}^5 \Delta t_i$ is determined by the three sweep times $\Delta t_1$, $\Delta t_3$ and $\Delta t_5$. For simplicity, we will assume in our discussions that the sweep times are equal, i.e., $t_s=\Delta t_1 = \Delta t_3 = \Delta t_5$.

The speed of the protocol for a given fidelity should be maximized for efficient use in quantum information processing. In general, a good initialization protocol should have $ \kappa_{\rm q}^{\rm I}\tau \ll 1$. This way one can perform multiple initializations during a coherent quantum computation. In our protocol, this requires the minimization of the combined duration of the three sweeps of the control parameter $\omega_{\rm L}(t)$ and the relaxation intervals, during which the actual initialization takes place. The length of the relaxation intervals is set by the relaxation rates and the desired fidelity, implying that after they are optimized, the duration of the protocol can be shortened only by faster sweeping. However, it is well known that fast changes in parameters can induce non-adiabatic transitions between the instantaneous eigenstates of the system~\cite{Landau32,*Zener32,*Stueckelberg32,*Majorana32}. Our requirements for the optimal operation points ($\omega_{\rm L}\approx \omega_{\pm}$) guarantee that our protocol is robust with respect to changes between the relative occupations of the instantaneous eigenstates during the first two sweeps. 
After the second relaxation process, the system lies in the ground state within the desired error $\alpha$. Thus, the essential sweep is the last one starting from the qubit resonance ($\omega_{\rm L}(t_4)=\omega_-$) and the ground state $|\Psi_0(t_4)\rangle$, and ending to the far off-resonant ground state $|\Psi_0(\tau)\rangle$. The relevant question is the following: how much of the ground state is excited during the final fast sweep? 

We can calculate the transition probabilities $P_{mn}\equiv P_{m\rightarrow n}(\tau;t_4) = |\langle \Psi_n(\tau)|\Psi_m(t_4)\rangle|^2$ between the low-energy eigenstates in the sudden approximation using the RWA results obtained in the previous section. However, since we are interested in the ground-state sweep fidelity $P_{\rm S}=P_{00}$ and since in the RWA the ground state is unaffected by the change of parameters, we have to include the contributions arising from the counter-rotating terms. 
In \hyperref[sec:Methods]{Methods}, we derive the worst-case estimate for the ground-state sweep fidelity in the sudden approximation. We obtain
\begin{equation}\label{eq:answeepfid}
P_{\rm S} = |\langle \Psi_0(\tau)|\Psi_0(t_4)\rangle|^2\approx 1-\left[\frac{g_{\rm LR}(\tau)}{\omega_{\rm L}(\tau)+\omega_{\rm R}}-\frac{g_{\rm LR}(t_4)}{\omega_{\rm q}+\omega_{\rm R}}\right]^2,
\end{equation}
where we assume that the sweep duration $\Delta t_5 \rightarrow 0$. Thus, the deviation from the perfect sweep fidelity is given by the difference between the perturbation parameters $g_{\rm LR}(t)/[\omega_{\rm L}(t)+\omega_{\rm R}]$ before and after the sweep. Order-of-magnitude estimates for typical superconducting circuit parameters give $g_{\rm LR}/(2\pi)\sim$ 100 MHz, $\omega_{\rm R}/(2\pi),\omega_{\rm q}/(2\pi)\sim$ 10 GHz and, accordingly, we have that the deviation from full fidelity is $1-P_{\rm S} \sim [g_{\rm LR}/(\omega_{\rm q}+\omega_{\rm R})]^2\sim10^{-4}$. 
Typically, the deviation is a couple of orders of magnitude smaller since the difference between the perturbation parameters is very small. We can therefore assume in our analytic considerations that the ground state is well protected during the sweeps of parameters in the Hamiltonian, and that the protocol duration can be estimated solely in terms of the relaxation rates of the static stages of the protocol, i.e. $\tau \approx \Delta t_2+\Delta t_4$. In the experimental implementation of the protocol, any cross coupling between the qubit and the flux line, used to adjust the left-resonator frequency, should be considered together with excitations of the SQUID. However, these issues seem not to significantly affect the achievable speed of our initialization protocol. 

\begin{figure}[t]
\centering
\includegraphics[width=0.7\linewidth]{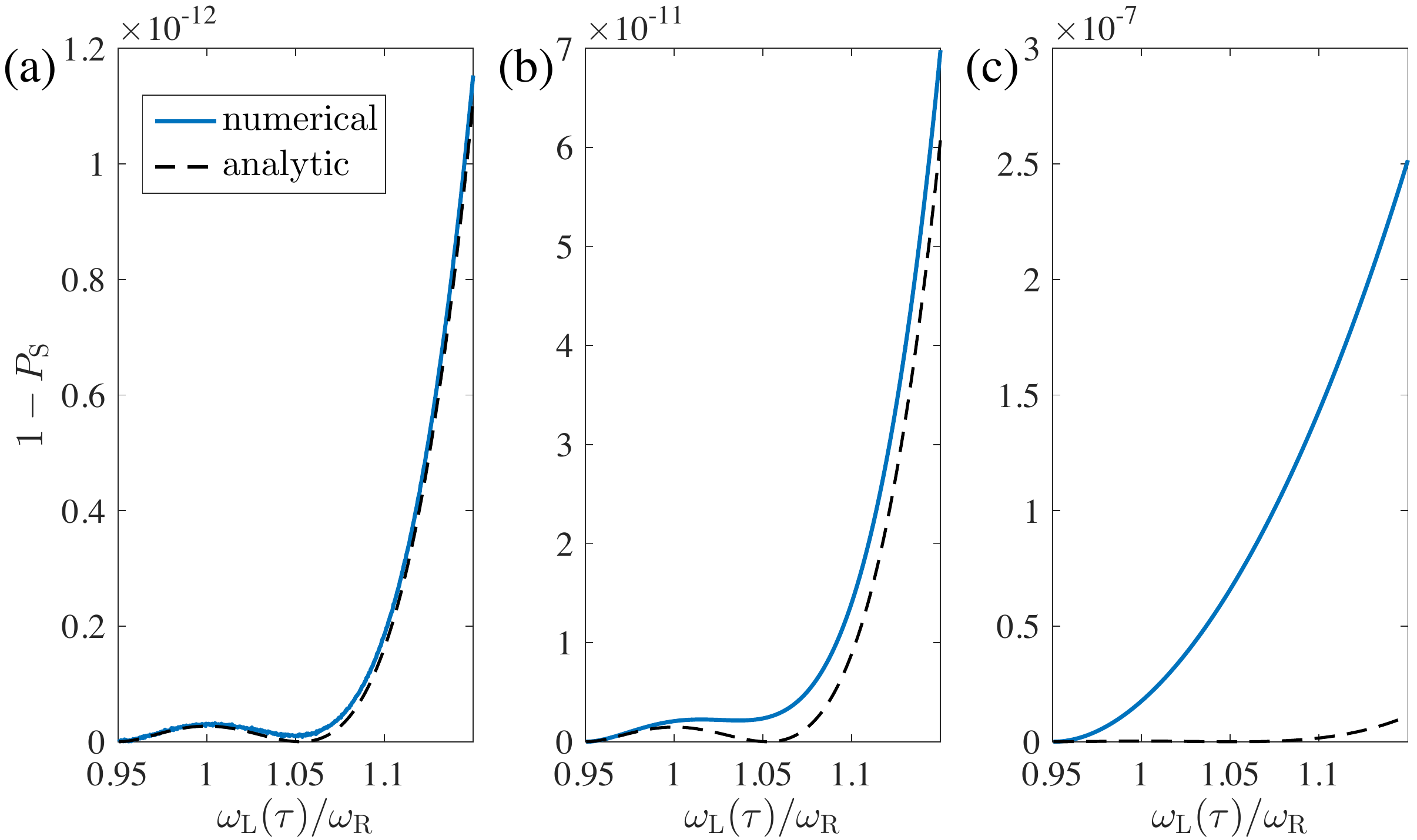}
\caption{Ground-state sudden sweep fidelity $1-P_{\rm S}$ as a function of the control parameter value $\omega_{\rm L}(\tau)$ at the end of the protocol.  We compare the analytically obtained sweep fidelity in Equation~(\ref{eq:answeepfid}) with that obtained by finding the ground state $|\Psi_0(t)\rangle$ numerically. We show the data for (a) $g_{\rm LR}^0 = 0.001\omega_{\rm R}$; (b) $g_{\rm LR}^0 = 0.0074 \omega_{\rm R}$ (value used in the simulations in Figs.~\ref{fig:stream},~\ref{fig:dynpops}, and~\ref{fig:fidvstau}); and (c) $g_{\rm LR}=0.1\omega_{\rm R}$. The qubit frequency is $\omega_{\rm q}=0.95\omega_{\rm R}$.}
\label{fig:gssfid}
\end{figure}

In Figure~\ref{fig:gssfid}, we compare the analytically obtained sweep fidelity from Equation~(\ref{eq:answeepfid}) with the sweep fidelity obtained by solving the instantaneous ground state $|\Psi_0(t)\rangle$ numerically. We observe that if the RWA is valid, i.e., for $g_{\rm LR}^0\ll \omega_{\rm R}$, the analytic result closely follows the numerical solution. For increasingly strong coupling, the second-order perturbation theory becomes insufficient which is visible as a large deviation between the analytic and numerical results. Even for $g_{\rm LR}^0=0.1\omega_{\rm R}$, however, the deviation from the perfect fidelity is of the order of $10^{-7}$, which indicates that the effects of the fast sweep  on the ground-state fidelity of the protocol can be neglected.

\subsection*{Markovian master equation for the time-dependent system}

In order to obtain more quantitative understanding of the protocol, we study its dynamics with the help of a Markovian master equation for the reduced system density operator $\hat{\rho}_{\rm S}(t) = \textrm{Tr}_{\rm E}\{\hat{\rho}(t)\}$, where the total density operator $\hat{\rho}(t)$ obeys the von Neumann equation
\begin{equation}
i\hbar\frac{\textrm{d}\hat{\rho}}{\textrm{d}t}=[\hat{H}(t),\hat{\rho}],
\end{equation}
where $\hat{H}(t)$ is defined in Equation~(\ref{eq:totalHam}). We first diagonalize the system Hamiltonian $\hat{H}_{\rm S}(t)$, as defined in Equation~(\ref{eq:HS}), in a time-independent basis $\{|n\rangle\}$ with the time-dependent unitary transformation $\hat{D}(t)=\sum_n |\Psi_n(t)\rangle\langle n|$. 
After the transformation, the time-evolution of $\hat{\rho}'(t) = \hat{D}^{\dag}(t)\hat{\rho}(t)\hat{D}(t)$ is governed by the effective system Hamiltonian
\begin{equation}
\hat{H}_{\rm eff}(t) = \hbar\sum_{nm}\left[\omega_{n}(t)\delta_{nm}-i\langle \Psi_n(t)|\dot{\Psi}_m(t)\rangle\right]|n\rangle\langle m|,
\end{equation}
where the latter term causes non-adiabatic transitions. Such term always appears if one wishes to preserve the form of the von Neumann equation in a time-dependent unitary transformation. After the transformation, the derivation of the master equation proceeds in a conventional manner~\cite{BreuerPetruccione,Scala07,Beaudoin11}: We assume that the 
initial state of the total system is uncorrelated, i.e. $\hat{\rho}(0)=\hat{\rho}_{\rm S}(0)\otimes\hat{\rho}_{\rm E}(0)$, and that the bath is in a thermal state, described by $\hat{\rho}_{\rm E}(0)$, throughout the temporal evolution. We consider only weak coupling to the environment and apply the standard Born and Markov approximations in the interaction picture, and subsequently trace over the environmental degrees of freedom. In the Markov approximation, the correlation time of the environment is negligibly short and we can express the master equation for $\hat{\rho}_{\rm S}'(t)$ in the secular approximation as
\begin{eqnarray}\label{eq:mastereq}
\frac{\textrm{d}\hat{\rho}_{\rm S}'(t)}{\textrm{d}t}&=&-\frac{i}{\hbar}[\hat{H}_{\rm eff}(t),\hat{\rho}_{\rm S}'(t)]\nonumber\\
&&+ \frac12\sum_{\omega_{nm}>0} \Gamma_{mn}(t)\left[2\hat{\pi}_{nm}\hat{\rho}_{\rm S}'(t)\hat{\pi}_{nm}^{\dag}- \hat{\pi}_{nm}^{\dag}\hat{\pi}_{nm}\hat{\rho}_{\rm S}'(t) -\hat{\rho}_{\rm S}'(t)\hat{\pi}_{nm}^{\dag}\hat{\pi}_{nm} \right]\nonumber\\
&&+\frac12\sum_{\omega_{nm}>0} \Gamma_{nm}(t)\left[2\hat{\pi}_{nm}^{\dag}\hat{\rho}_{\rm S}'(t)\hat{\pi}_{nm} -\hat{\pi}_{nm}\hat{\pi}_{nm}^{\dag}\hat{\rho}_{\rm S}'(t) -\hat{\rho}_{\rm S}'(t)\hat{\pi}_{nm}\hat{\pi}_{nm}^{\dag}\right]\nonumber\\ 
&&+\frac12\sum_{n}\Gamma_{nn}(t)\left[2\hat{\pi}_{nn}\hat{\rho}_{\rm S}'(t)\hat{\pi}_{nn}-\hat{\pi}_{nn}\hat{\pi}_{nn}\hat{\rho}_{\rm S}'(t)-\hat{\rho}_{\rm S}'(t)\hat{\pi}_{nn}\hat{\pi}_{nn}\right].
\end{eqnarray}
Above, we have defined the ladder operators between the static basis states as $\hat{\pi}_{nm}=|n\rangle\langle m|$. The instantaneous transition rates $\Gamma_{mn}(t)$ are identical to those obtained with Fermi's golden rule in Equation~(\ref{eq:statrate})~\cite{Alicki77}. Similar to the case of a static Hamiltonian~\cite{BreuerPetruccione,Scala07}, the environmental decoherence is included in the Lindblad terms on the last three rows of the master equation. The first term models the transitions between the adiabatic states that dissipate energy to the environment, the second term represents absorption from the environment and the last term induces dephasing of the adiabatic states. We note that the secular approximation made above is justified if the relaxation rates are small compared to the minimum separation between the eigenfrequencies of the system~\cite{BreuerPetruccione}, i.e., we have
\begin{equation}\label{eq:seccond}
\Gamma_{mn}(t)\ll \min_{i\neq j}|\omega_i(t)-\omega_j(t)|.
\end{equation} 

We note that the adiabatic master equation above can be improved by making successive diagonalizing transformations for $\hat{H}_{\rm eff}(t)$. As a result, the master equation is represented in the basis of the so-called superadiabatic states, the time-dependence of which is typically suppressed after each diagonalizing transformation. Such adiabatic renormalization was first described for a general time-dependent quantum system by Berry~\cite{Berry87}, and later applied to studies of dissipation in driven superconducting qubits~\cite{Salmilehto10,Salmilehto11,Suomela15}. The lowest-order superadiabatic correction was studied in references~\onlinecite{Pekola10,Solinas10}.


\subsection*{Protocol speed}

Let us study the decay of an excitation in our system by solving the master equation~(\ref{eq:mastereq}) for an initial state spanned by the low-energy adiabatic states as $|\Psi(0)\rangle = \sum_{n=0}^4 a_n(0)|\Psi_n(0)\rangle$. We assume that the environment is so cold that thermal excitations of the system are negligible. This guarantees that the quantum state $\hat \rho_{\rm S}(t)$ of the system remains in the low-energy subspace during the temporal evolution. In the beginning of a realistic initialization procedure, we may have no knowledge on the distribution of the occupations $P_n(0)=|a_n(0)|^2$. Therefore, we choose the operation points of our protocol such that
\begin{equation}\label{eq:apprrates1}
t\in [t_1,t_2]: \ \Gamma_2\equiv \Gamma_{20}(t)=\Gamma_{30}(t)=\Gamma_0/2, \ \Gamma_{10}(t)\approx 0,
\end{equation}
and
\begin{equation}\label{eq:apprrates2}
t\in[t_3,t_4]: \  \Gamma_1\equiv \Gamma_{10}(t)=\Gamma_{20}(t)=(\omega_{\rm q}/\omega_{\rm R})^2\Gamma_0/2, \ \Gamma_{30}(t)\approx 0,
\end{equation}
which is the case for $\omega_{\rm L}(t_2)\approx \omega_+ \approx \omega_{\rm R} + g_{\rm Rq}^2/\delta_{\rm Rq}$ and $\omega_{\rm L}(t_4)\approx \omega_- \approx \omega_{\rm q} - g_{\rm Rq}^2/\delta_{\rm Rq}$, respectively. 

Before discussing the numerical results, we present a simple analytic estimate for the initialization fidelity. The general form for the excited-state occupations can be solved from the master equation~(\ref{eq:mastereq}) and written as
\begin{equation}\label{eq:expopDT}
P_i(t) = P_i(0)e^{-\int_0^t \Gamma_{i0}(t') dt'}.
\end{equation} 
By neglecting the small contributions of the fast sweeps, these can be readily written in terms of Equations~(\ref{eq:apprrates1}) and~(\ref{eq:apprrates2}). 
We find that the deviation from the perfect fidelity, $\alpha(\tau) = 1-P_0(\tau) = \sum_{i=1}^3 P_i(\tau)$, can be written as
\begin{equation}
\frac{\alpha(\tau)}{\alpha(0)} = \overline{P}_1(0)e^{-\Gamma_1 \Delta t_4} + \overline{P}_2(0)e^{-\Gamma_2 \Delta t_2} +\overline{P}_3(0)e^{-\Gamma_1 \Delta t_4-\Gamma_2 \Delta t_2},
\end{equation}
where the relative initial occupations of the excited states have been defined as $\overline{P}_i(0)=P_i(0)/\sum_{j=1}^3 P_j(0)$ for $i=1,2,3$. The tolerable deviation $\alpha(\tau)$ from perfect fidelity at the end of the protocol is fixed in the beginning of the protocol. 
Since we do not know the relative occupations in the beginning of the initialization protocol, we have to wait the times $\Delta t_2\approx\textrm{ln }[\alpha(0)/\alpha(\tau)]/\Gamma_2$ and $\Delta t_4\approx\textrm{ln }[\alpha(0)/\alpha(\tau)]/\Gamma_1$ so that any excitation in the system is decayed down to the desired accuracy. As a consequence, we obtain an upper bound for the total wait time $\overline{\tau}=\Delta t_2+\Delta t_4$ of the protocol as
\begin{equation}\label{eq:analyticTvsfid}
\overline{\tau} \leq \textrm{log}_{10}[\alpha(0)/\alpha(\tau)]\frac{2\textrm{ln}(10)}{\Gamma_0}\left[1+\left(\frac{\omega_R}{\omega_q}\right)^2\right] = \beta \tau_{10}.
\end{equation}
Above, we denote $\alpha(\tau)=10^{-\beta}\alpha(0)$ where $\beta \geq 0$, and $\tau_{10}=2\textrm{ln}(10)[1+(\omega_{\rm R}/\omega_{\rm q})^2]/\Gamma_0$. For $\omega_{\rm R}\gtrsim \omega_{\rm q}$, we have that $\tau_{10}\approx 4\textrm{ln}(10)/\Gamma_0$ which sets the time scale for the decrease of $\alpha(\tau)$ by an order of magnitude. We note that the secular approximation used in the derivation of the master equation requires through Equation~(\ref{eq:seccond}) that
\begin{equation}
\Gamma_0\ll \min \left\{g_{\rm LR}^{0}, \ g_{\rm LR}^0\frac{g_{\rm Rq}}{\delta_{\rm Rq}}\sqrt{\frac{\omega_{\rm q}}{\omega_{\rm R}}}\right\}=g_{\rm LR}^0\frac{g_{\rm Rq}}{\delta_{\rm Rq}}\sqrt{\frac{\omega_{\rm q}}{\omega_{\rm R}}}\equiv \Gamma_0^{\rm max},
\end{equation}
where the equality holds since in our case $0<g_{\rm Rq}\ll \delta_{\rm Rq}$ and $\omega_{\rm q}<\omega_{\rm R}$. Additionally, we required that in the off state of the protocol, the intrinsic rates dominate over those of the engineered environment, i.e., $\Gamma_{10}(0)= \alpha_{\rm q}\kappa_{\rm q}^{\rm I}$ and $\Gamma_{20}(0)= \alpha_{\rm R}\kappa_{\rm R}^{\rm I}$, where $\alpha_{\rm q},\alpha_{\rm R}\ll 1$. We show in \hyperref[sec:Methods]{Methods} that these lead to the condition $\Gamma_0\leq\omega_{\rm R}\sqrt[3]{\gamma^2\alpha_{\rm q}\kappa_{\rm q}^{\rm I}\delta_{\rm Lq}^2(0)/\{[\delta_{\rm Lq}(0)+\omega_{\rm q}]^2\omega_{\rm R}\}}$, where $\gamma = \Gamma_0/\Gamma_0^{\rm max}\ll 1$ and $\delta_{\rm Lq}(0) = \omega_{\rm L}-\omega_{\rm q}$ is the detuning between the left-resonator and the qubit in the off state of the protocol. 
This sets a lower bound for the decay time:
\begin{equation}
\tau_{10}\geq \frac{4\ln(10)}{\omega_{\rm R}}\sqrt[3]{\frac{[\delta_{\rm Lq}(0)+\omega_{\rm q}]^2\omega_{\rm R}}{\gamma^2\alpha_{\rm q}\kappa_{\rm q}^{\rm I}\delta_{\rm Lq}^2(0)}} 
= 300 \ \textrm{ns},
\end{equation}
where the numerical estimate is made for typical superconducting circuit parameters $\omega_{\rm R}/(2\pi)= 10$ GHz, $\omega_{\rm q}/(2\pi) = 9.5$ GHz, $\delta_{\rm Lq}(0)/(2\pi) = 2$ GHz, and $\kappa_{\rm q}^{\rm I} =  10^4$ s$^{-1}$. We have used $\beta = 0.5$ and confirmed with a classical calculation of the transition rate $\Gamma_2$ that the secular approximation holds within a relative error of $4\%$. We also set $\alpha_{\rm q}=0.1$. The lower bound for the decay time above would represent a significant improvement to that of the current experimental benchmark for qubit ground-state initialization protocol~\cite{Geerlings13}. By choosing the off state infinitely far from the qubit ($\delta_{\rm Lq}\rightarrow \infty$), the decay time can be improved to $\tau_{10}\gtrsim 90$ ns. In the following, we set $\alpha(0)=1$ in order to obtain an estimate for the total wait time for initialization of a qubit excitation. For example, if one wishes to obtain a ground state fidelity of $\alpha(\tau) = 10^{-3}$ with our protocol, the wait time should be of the order of $\overline{\tau} = 3\tau_{10}$. 

\begin{figure}[t]
\centering
\includegraphics[width=0.7\linewidth]{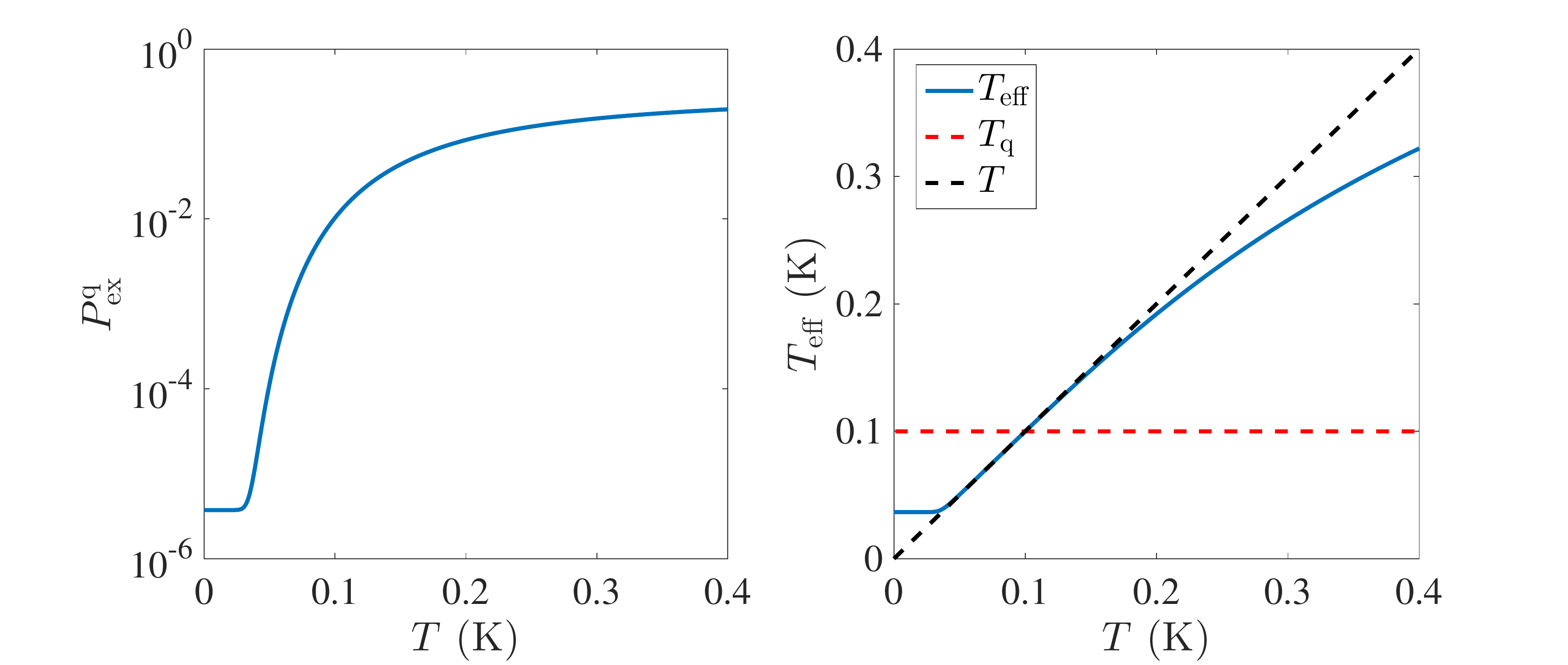}
\caption{
(a) Analytic excited state occupation $P_{\rm ex}^{\rm q}$; and (b) the corresponding effective qubit temperature $T_{\rm eff}$ after the initialization protocol as functions of the resistor temperature $T$, cf. Equation~(\ref{eq:Pexq}). In (b), we also plot the temperatures for the intrinsic qubit (red dashed line) the resistive bath (black dashed line). We use the intrinsic qubit temperature $T_{\rm q}=100$ mK and the intrinsic relaxation rate $\kappa_{\rm q}^{\rm I}= 10^4$ s$^{-1}$. Other parameters are identical to those in Figure~\ref{fig:stream}. }
\label{fig:Teff}
\end{figure}

\subsection*{Qubit temperature}

Above, we calculated the deviation $\alpha(\tau)$ from the perfect fidelity by neglecting the intrinsic dissipation in the qubit. However, in addition to the temperature $T$ of the resistor, the fidelity is reduced by the intrinsic temperature $T_{\rm q}$ of the qubit. We estimate this effect by first defining the effective qubit temperature $T_{\rm eff}$ in terms of the excited-state occupation $P_{\rm ex}^{\rm q}$ of the qubit at the end of the protocol as
\begin{equation}
P_{\rm ex}^{\rm q} \equiv \frac{1}{1+e^{\hbar \omega_{\rm q}/(k_{\rm B} T_{\rm eff})}}.
\end{equation}
By using the principle of detailed balance, we can solve $P_{\rm ex}^{\rm q}$ in our four-state model from
\begin{equation}
P_0\left[\Gamma_{01}+\Gamma_{02} +\kappa_{\rm q}^{\rm I}e^{-\hbar\omega_{\rm q}/(k_{\rm B}T_{\rm q})}\right] = P_1\Gamma_{10} + P_2\Gamma_{20} + P_{\rm ex}^{\rm q}\kappa_{\rm q}^{\rm I},
\end{equation}
where $P_{\rm ex}^{\rm q}=\frac12(P_1+P_2)$ 
and we have included the coupling $\kappa_{\rm q}^{\rm I}$ between the qubit and its intrinsic environment. The occupation at the end of the protocol can be calculated at the second operation point, where we use $\Gamma_{10}=\Gamma_{20} = (\omega_{\rm q}/\omega_{\rm R})^2\Gamma_0/2 = \Gamma_1$, $\Gamma_{30}=0$, and the detailed balance relation $\Gamma_{0i} = \Gamma_{1}\exp[-\hbar\omega_{0i}/(k_{\rm B}T)]$ for the excitation and absorption rates. Above, we have neglected the intrinsic excitation of the right resonator. We note that with the above assumptions $P_3 =0$ and, thus, $P_0 = 1-2P_{\rm ex}^{\rm q}$. Thus, we can write the excited state population for the qubit as
\begin{equation}\label{eq:Pexq}
P_{\rm ex}^{\rm q} =  \frac{e^{-\hbar\omega_1^{\downarrow}/(k_{\rm B}T)}+e^{-\hbar\omega_2/(k_{\rm B}T)} + (\kappa_{\rm q}^{\rm I}/\Gamma_1)e^{-\hbar\omega_{\rm q}/(k_{\rm B}T_{\rm q})}}{2+\kappa_{\rm q}^{\rm I}/\Gamma_1 + 2\left[e^{-\hbar\omega_1^{\downarrow}/(k_{\rm B}T)}+e^{-\hbar\omega_2/(k_{\rm B}T)}+(\kappa_{\rm q}^{\rm I}/\Gamma_1)e^{-\hbar\omega_{\rm q}/(k_{\rm B}T_{\rm q})}\right]}= \frac{1}{1+e^{\hbar\omega_{\rm q}/(k_{\rm B}T_{\rm eff})}}.
\end{equation}
We show in Figure~\ref{fig:Teff} the above analytic excited qubit state occupation $P_{\rm ex}^{\rm q}$ and the corresponding effective temperature $T_{\rm eff}$. For our example parameters (see \hyperref[sec:Methods]{Methods}) and for the intrinsic qubit temperature $T_{\rm q}=100$ mK, the data show a saturation of the qubit excitation and the effective qubit temperature to values $P_{\rm ex}^{\rm q} \approx 4\times 10^{-6}$ and $T_{\rm eff} \approx 36$ mK, respectively, for $T\lesssim 30$ mK. In our simulations, we use the value $T=10$~mK. In the remaining calculations, we have neglected the intrinsic qubit dissipation in order to keep our results independent on the temperature of the intrinsic environment of the qubit.

\subsection*{Numerical results}

\begin{figure}[t]
\centering
\includegraphics[width=\linewidth]{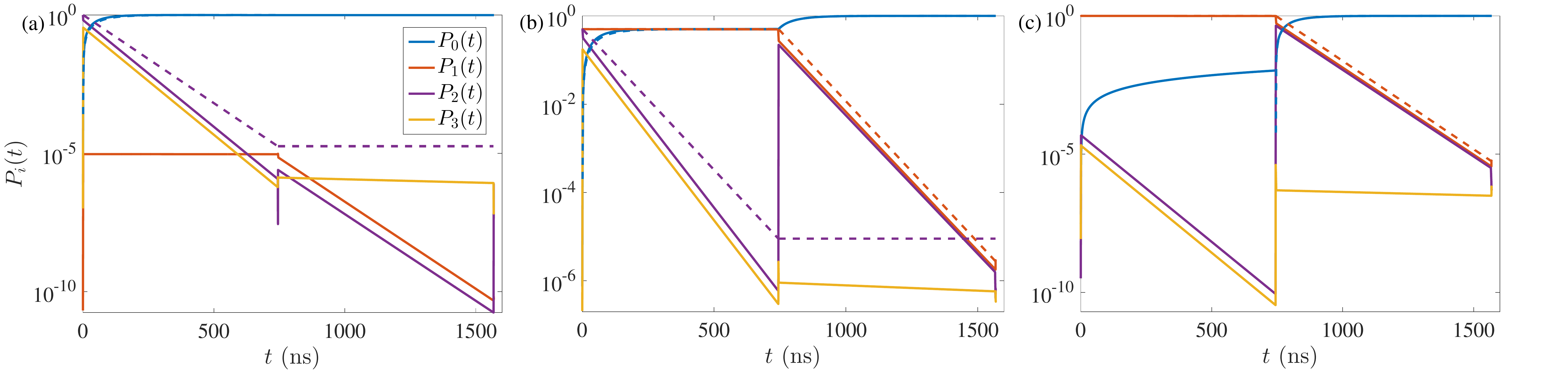}
\caption{Dynamics of the occupation probabilities of the low-energy eigenstates. We show results for the initial probabilities $P_0(0)=P_3(0)=0$, and (a) $P_1(0) = 0$ and $P_2(0)=1$; (b) $P_1(0)=P_2(0)=0.5$; (c) $P_1(0)=1$ and $P_2(0)=0$. Analytic occupation probabilities are shown with dashed lines, and obtained from Equation~(\ref{eq:expopDT}) by assuming sudden sweeps. 
We use parameters identical to those in Figure~\ref{fig:stream}. 
In the numerical solutions, the sweep time $t_{\rm s} = 1$ ns and $\alpha(\tau) = 10^{-5}$.}
\label{fig:dynpops}
\end{figure}

\begin{figure}[t]
\centering
\includegraphics[width=0.6\linewidth]{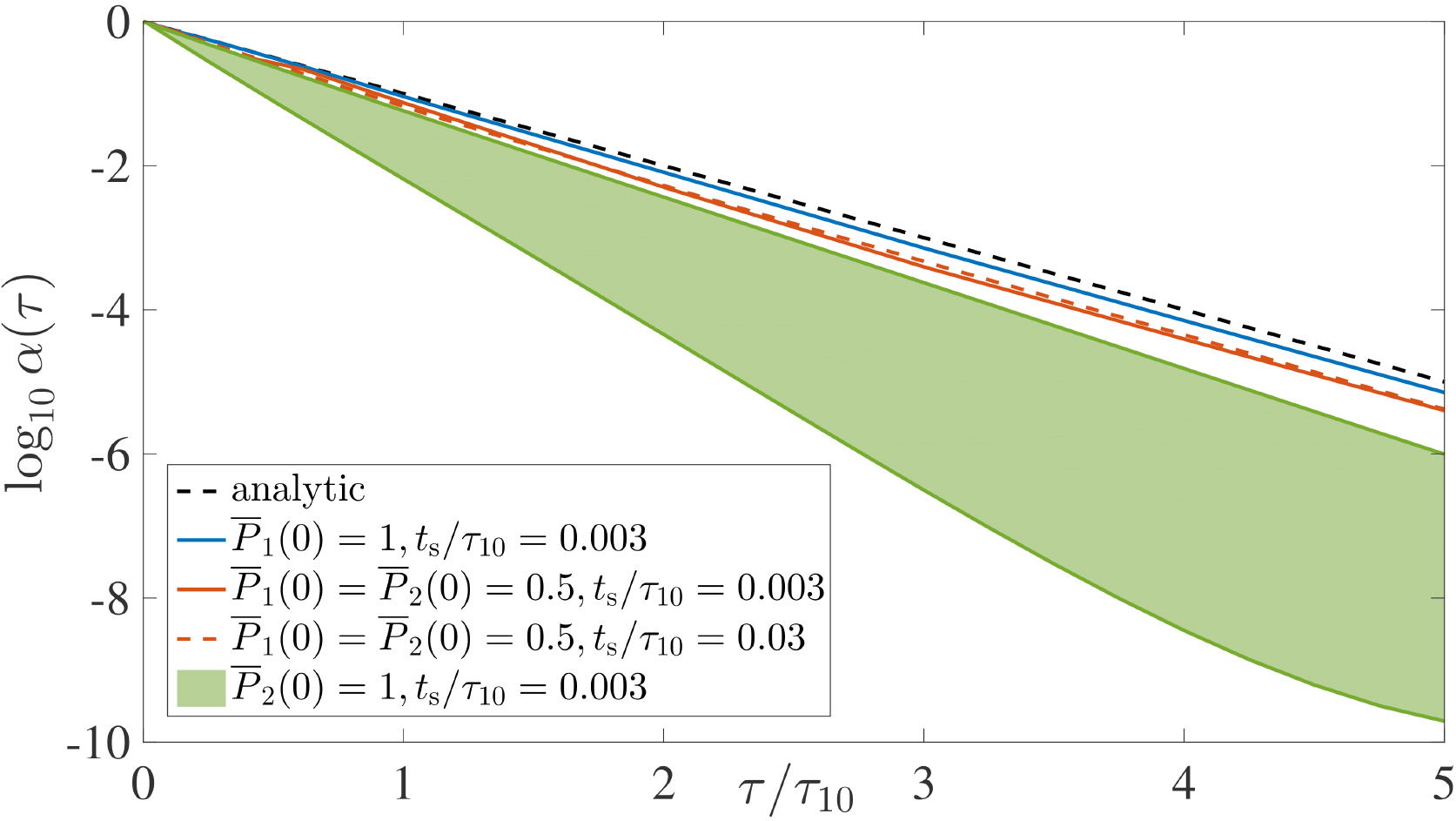}
\caption{Qubit initialization error as a function of the protocol duration. We present a comparison between different initial states and sweep times. For $\overline{P}_2(0)=1$, the initialization error is a strongly oscillating function of the protocol duration and its values are located within the shaded region. The physical parameters are identical to those in Figure~\ref{fig:stream}. The decay time $\tau_{10}=300$ ns.}
\label{fig:fidvstau}
\end{figure}

Let us compare the analytic model above with the numerical solution of the master equation~(\ref{eq:mastereq}). To this end, we solve the master equation by truncating to the subspace spanned by the five lowest adiabatic energy eigenstates 
which are obtained by diagonalizing the instantaneous Hamiltonian $\hat{H}_{\rm S}(t)$, defined in Equation~(\ref{eq:HS}). We have confirmed that the relative errors caused by the truncation in the ground-state occupation are of the order of $10^{-7}$ or smaller. In Figure~\ref{fig:dynpops}, we present the dynamics of the occupations $P_i(t)$ for the four lowest-energy states. 
We study the decay of a single excitation with three different initial occupation probabilities. 
Our choice for the off-state (see Figure~\ref{fig:stream}) 
guarantees that the three degrees of freedom are initially weakly coupled and the states $|\Psi_1(0)\rangle$, $|\Psi_2(0)\rangle$, and $|\Psi_3(0)\rangle$ can be well approximated by the first excited states of the qubit, the right resonator, and the left resonator, respectively. 
In the numerical simulations, we choose the 
wait times at the operation points equal to those presented in the analytic model. If the excitation belongs initially to the qubit, the numerical results are in very good agreement with the analytic occupations obtained in Equation~(\ref{eq:expopDT}). During the sweep to the qubit resonance, part of the initial qubit occupation is transferred non-adiabatically to the left resonator. However, the final numerical fidelity is close to the analytic estimate since at the second operation point the qubit and the left resonator become hybridized and the relaxation rates for the resulting two states are equal. This is clearly visible in Figure~\ref{fig:fidvstau} where we show the dependence of the protocol error $\alpha(\tau)$ on the total duration $\tau$. Similarly, if the right resonator has a non-vanishing initial occupation, part of it is transferred to the left-resonator during the first sweep. Again, this does not lead to major deviations from the analytic fidelity due to the equal relaxation rates at the first operating point. 

We note that one has to be careful when tuning to the first operation point. If the operation point is above the resonance, i.e., $\omega_{\rm L}(t_1)>\omega_+$, any occupation remaining in the state $|\Psi_2(t)\rangle$ at $t=t_2$ will be transferred to the state $|\Psi_3(t)\rangle$ in a Landau--Zener-type process when the system is subsequently swept across the avoided crossing. This leads to a decrease in the protocol fidelity, since at the second operation point the relaxation rate $\Gamma_{30} \approx 10^{-2}\Gamma_0$ (see Figure~\ref{fig:stream}). 

If the desired fidelity $P_0(\tau)=1-\alpha(\tau)$ is close to unity, the wait times have to be long and, as a consequence, the choice for the initial location of the excitation does not have a large influence on the protocol time. The analytic estimate~(\ref{eq:analyticTvsfid}) serves as an upper bound for the total wait time $\overline{\tau}$, and we approach the upper bound in the case of the full qubit excitation. Fortunately, the numerically obtained fidelity is always higher than that given by the analytic upper bound for the wait time, since there always exist some residual relaxation to ground state from the nonresonant states. If the excitation is partly or completely in the right-resonator, the desired fidelity is reached faster than in the analytic prediction, as depicted in Figure~\ref{fig:fidvstau}. 
This occurs because the part of the initial right-resonator occupation that has not decayed at the first operation point can continue the decay at the second operation point due to the hybridization with the left resonator which causes occupation transfer during the first sweep. We also observe that for our choice of parameters the relatively strong coupling between the qubit and the right resonator, $g_{\rm Rq}/\delta_{\rm Rq}\approx 0.14$, causes oscillations in the protocol fidelity as a function of the protocol time $\tau$. 

\section*{Discussion}

In summary, we have proposed and modeled a qubit initialization protocol where the coupling between a superconducting qubit and an engineered environment can be externally controlled with a tunable resonator. Using experimentally realizable parameters, we have solved the time-dependent Markovian master equation for the protocol and shown that the tunable resonator can be used for fast and precise reset of a qubit excitation. We have also demonstrated that fast changes of the bare angular frequency of the tunable resonator do not reduce the final ground-state fidelity of the protocol. As a result, the duration of the protocol for a given fidelity can be estimated in terms of the decay rate of the tunable resonator, $\Gamma_0$, with a simple analytic model. We also found that the present experimental state-of-the-art decay time~\cite{Geerlings13} for ground-state qubit initialization may be decreased almost by an order of magnitude. Moreover, we observed that at dilution refridgerator temperatures ($T\lesssim 30$ mK) the effective qubit temperature can be reduced to one third of its intrinsic temperature of 100 mK, resulting in an excited qubit state occupation of roughly $10^{-6}$. 

In the off-mode of the protocol, the coupling to the engineered environment has to be weak enough such that the decoherence during quantum computation is determined mainly by the intrinsic environment of the qubit. This sets a limitation for the protocol speed since the decay rate $\Gamma_0$ has to be lower than the intrinsic dissipation rates in the system. Furthermore, the assumption of weak coupling with the environment restricts the possible values of decay rates. 
In future work, the evolution of the reduced density operator should be solved in the regime of strong environmental coupling which may lead to further improvements in the protocol speed. Also, the possibilities of combining our protocol with driven reset method~\cite{Geerlings13} should be investigated for further improvements on the protocol duration and speed.




\section*{\label{sec:Methods}Methods} 

\subsection*{Jacobian diagonalization}

We perturbatively solve the eigenproblem of the instantaneous Hamiltonian~(\ref{eq:HS}).
In the RWA, the occupation number $\hat{N}\equiv \hat{a}^{\dag}_{\rm L}\hat{a}_{\rm L} + \hat{a}^{\dag}_{\rm R}\hat{a}_{\rm R} + \hat{\sigma}_+\hat{\sigma}_-$ is a conserved quantity because $[\hat{H}_S,\hat{N}]=0$. This means that $\hat{H}_{\rm S}(t)$ and $\hat{N}$ have joint eigenstates. Eigenvalues of the occupation number are $N=n_{\rm L}+n_{\rm R} + n_{\rm q}=0,1,2,\ldots$, where $n_{\rm L}$ and $n_{\rm R}$ are the occupation numbers of the left and right resonators, respectively, and assume values $n_{\rm L},n_{\rm R} = 0,1,2,\ldots$. The number $n_{\rm q}$ denotes the qubit occupation and assumes values 0 and 1. Each occupation number $N$ belongs to $(2N+1)$ degenerate eigenstates of the form $|n_{\rm L},n_{\rm R},n_{\rm q}\rangle$. These are also the eigenstates of the unperturbed Hamiltonian $\hat{H}_0(t)$. We note that if the counter-rotating terms neglected in the RWA are included, the occupation number is no longer conserved and the above arguments do not hold. In the numerical simulations we employ the non-RWA Hamiltonian.

We are interested in the transitions between the low-energy eigenstates. The $N=1$ manifold consists of three states: $\{|0,0,{\rm e}\rangle, |0,1,{\rm g}\rangle, |1,0,{\rm g}\rangle\}$ where, for clarity, we use the symbols g $ \leftrightarrow 0$ and e $ \leftrightarrow 1$ for the qubit occupation number $n_{\rm q}$ in the ground and excited state, respectively. As a consequence, the eigenstates of the RWA Hamiltonian with occupation number $N=1$ are linear combinations of these three basis states. 
Therefore, it is enough to find eigenstates for the truncated $3\times3$ Hamiltonian matrix
\begin{equation}
H_{\rm S}(t) = \hbar\left(\begin{array}{ccc}
\omega_{\rm q} & -ig_{\rm Rq}& 0\\
ig_{\rm Rq} & \omega_{\rm R} & g_{\rm LR}(t)\\
0 & g_{\rm LR}(t) & \omega_{\rm L}(t)
\end{array}\right).
\end{equation}

We rely on approximative methods for an intuitive analytic solution for an arbitrary value of $\omega_{\rm L}(t)$. The challenge is to find the solution for nearly resonant states, 
for which non-degenerate perturbation theory fails. The coupling between the left-resonator and the qubit is of second order in the coupling frequencies $g_{\rm LR}^0$ and $g_{\rm Rq}$, which causes slow convergence in the conventional nearly degenerate and Brillouin--Wigner perturbation theories. Instead, we employ the Jacobian diagonalization~\cite{Jacobi1846,NR11} 
for the eigenproblem of the $N=1$ manifold. We are especially interested in the transition rates when the left resonator is in resonance either with the right resonator or the qubit. 

The Jacobian diagonalization reduces a hermitian matrix into a diagonal form by a sequence of rotations on its two-dimensional submatrices as follows: We first make a rotation that diagonalizes $H_{\rm S}(t)$ in the subspace spanned by $|0,1,{\rm g}\rangle$ and $|0,0,{\rm e}\rangle$, since then the tuned left resonator is effectively coupled to the other two degrees of freedom, and also because in our system $g_{\rm Rq}\gg g_{\rm LR}(t)$ which guarantees rapid convergence in the diagonalization method based on the Jacobian transformations. We obtain
\begin{equation}\label{eq:Jac1}
H_{\rm S}(t) = \hbar\left(\begin{array}{ccc}
\omega_- & 0 & G_{\rm L-}(t)\\
0 & \omega_{+} & G_{\rm L+}(t)\\
G_{\rm L-}(t) & G_{\rm L+}(t) & \omega_{\rm L}(t)
\end{array}\right).
\end{equation}
Above, we have defined $\omega_{\pm}=\omega_{\rm Rq}^{\rm av}\pm\Omega_{\rm Rq}$, $\omega_{\rm Rq}^{\rm av}=(\omega_{\rm R}+\omega_{\rm q})/2$, $\Omega_{\rm Rq}=\sqrt{(\delta_{\rm Rq}/2)^2+g_{\rm Rq}^2}$, $\delta_{\rm Rq}=\omega_{\rm R}-\omega_{\rm q}$, and $G_{\rm L\pm}(t)\equiv \langle 1,0,\textrm{g}|\hat H_{\rm S}(t)|\pm\rangle=g_{\rm LR}(t)\sqrt{[1\pm\delta_{\rm Rq}/(2\Omega_{\rm Rq})]/2}$. In Equation~(\ref{eq:Jac1}), Hamiltonian $\hat{H}_{\rm S}$ is written in the basis spanned by $\{|-\rangle,|+\rangle,|1,0,{\rm g}\rangle\}$, where
\begin{eqnarray}
|\pm\rangle &=&\frac{1}{\sqrt{2}}\left[\sqrt{1\pm\delta_{\rm Rq}/(2\Omega_{\rm Rq})}|0,1,{\rm g}\rangle \mp i\sqrt{1\mp\delta_{\rm Rq}/(2\Omega_{\rm Rq})}|0,0,{\rm e}\rangle\right].
\end{eqnarray}
At this point, the left resonator is coupled to the hybridized qubit and right resonator states $|\pm\rangle$, which in the dispersive regime ($g_{\rm Rq}\ll|\delta_{\rm Rq}|$) have the energies
\begin{equation}
\omega_{\pm}=\omega_{\rm Rq}^{\rm av}\pm \sqrt{(\delta_{\rm Rq}/2)^2+g_{\rm Rq}^2}\approx \left\{\begin{array}{c}
\omega_{\rm R} + \frac{g_{\rm Rq}^2}{\delta_{\rm Rq}};\\
\omega_{\rm q} - \frac{g_{\rm Rq}^2}{\delta_{\rm Rq}},
\end{array}\right.
\end{equation} 
where we have anticipated our choice of parameters and assumed that $\delta_{\rm Rq}>0$. If the left resonator is far detuned, $|\omega_{\rm L}(t)-\omega_{\pm}|\gg G_{\rm L\pm}$, the above basis states accurately approximate the true eigenstates. In the vicinity of resonances with $|\omega_{\rm L}(t)-\omega_{\pm}|\ll G_{\rm L\pm}$, one has to take effects arising from the coupling terms into account by making subsequent diagonalizations in the resonant subspaces. However, if $G_{\rm L\pm}\ll |\omega_+-\omega_-|$ we can neglect the contributions caused by the off-resonant coupling term $G_{\rm L \mp}$. 

Since $G_{\rm L+}>G_{\rm L-}$, we improve on the convergence by diagonalizing the Hamiltonian~(\ref{eq:Jac1}) in the subspace spanned by $|1,0,{\rm g}\rangle$ and $|+\rangle$, and obtain
\begin{equation}\label{eq:JacHam2}
H_{\rm S}(t) = \hbar\left(\begin{array}{ccc}
\omega_{-} & G_{\downarrow -}(t) & G_{\uparrow -}(t)\\
G_{\downarrow -}(t) & \omega_{\downarrow}(t) & 0\\
G_{\uparrow -}(t) & 0 & \omega_{\uparrow}(t)
\end{array}\right),
\end{equation}
where $\omega_{\uparrow\!/\!\downarrow}(t)=\omega_{\rm L+}^{\rm av}(t)\pm\Omega_{\rm L+}(t)$, $\omega_{\rm L+}^{\rm av}(t)=[\omega_{\rm L}(t) + \omega_+]/2$, $\Omega_{\rm L+}(t)=\sqrt{[\delta_{\rm L+}(t)/2]^2+G_{\rm L+}^2(t)}$, $\delta_{\rm L+}(t)=\omega_{\rm L}(t)-\omega_{+}$, and $G_{\uparrow\!/\!\downarrow -}(t)=G_{\rm L-}(t)\sqrt{\{1\pm \delta_{\rm L+}(t)/[2\Omega_{\rm L+}(t)]\}/2}$. Furthermore,
\begin{eqnarray}
|\!\uparrow\!/\!\downarrow\!(t)\rangle &=&\frac{1}{\sqrt{2}}\left[\sqrt{1\pm\delta_{\rm L+}(t)/[2\Omega_{\rm L+}(t)]}|1,0,{\rm g}\rangle \pm \sqrt{1\mp\delta_{\rm L+}(t)/[2\Omega_{\rm L+}(t)]}|+\rangle \right]
\end{eqnarray}
are the states corresponding to the approximative eigenvalues $\hbar\omega_{\uparrow/\downarrow}(t)$. Thus in Equation~\ref{eq:JacHam2}, the Hamiltonian $\hat{H}_{\rm S}(t)$ is represented in the basis $\{|-\rangle,|\!\downarrow\!(t)\rangle, |\!\uparrow\!(t)\rangle\}$.

If $G_{\rm L\pm}(t)\approx |\omega_+-\omega_-|$, we cannot neglect the off-resonant coupling even for nearly degenerate states. We also wish to remove the possible degeneracy between the left resonator and the qubit. We note that if $\omega_{\rm L}(t)<\omega_+$, the left-resonator excited state is approximately given by $|1,0,\textrm{g}\rangle \approx |\!\downarrow\!(t)\rangle$, and in the opposite limit $\omega_{\rm L}(t)> \omega_+$, we have $|1,0,\textrm{g}\rangle \approx |\!\uparrow\!(t)\rangle$. Therefore, we need to calculate the corrections for both states $|\!\uparrow\!/\!\downarrow\!(t)\rangle$ caused by the effective qubit state $|-\rangle$. Thus, we further diagonalize the Hamiltonian in the subspaces spanned by $\{|-\rangle, |\!\downarrow\!(t)\rangle\}$ and $\{|-\rangle, |\!\uparrow\!(t)\rangle\}$ and denote the results with $\downarrow$ and $\uparrow$, respectively. We obtain
\begin{equation}
H_{\rm S}(t) \approx \hbar\left(\begin{array}{ccc}
\omega_1^{\!\uparrow\!/\!\downarrow}(t) & 0 & 0\\
0 & \omega_2(t) & 0\\
0 & 0 & \omega_3(t)
\end{array}\right),
\end{equation}
where have neglected the small couplings between the new approximate eigenstates and defined $\omega_1^{\!\uparrow\!/\!\downarrow}(t)=\omega_{\,\!\uparrow\!/\!\downarrow -}^{\rm av}(t)-\Omega_{\,\!\uparrow\!/\!\downarrow -}(t)$, $\omega_2(t)=\omega_{\downarrow -}^{\rm av}(t)+\Omega_{\downarrow -}(t)$, $\omega_3(t)=\omega_{\uparrow -}^{\rm av}(t)+\Omega_{\uparrow -}(t)$, $\omega_{\,\!\uparrow\!/\!\downarrow -}^{\rm av}(t) = [\omega_{\,\!\uparrow\!/\!\downarrow}(t) + \omega_-]/2$, $\Omega_{\,\!\uparrow\!/\!\downarrow -}(t)=\sqrt{(\delta_{\,\!\uparrow\!/\!\downarrow -}(t)/2)^2+G_{\,\!\uparrow\!/\!\downarrow -}^2(t)}$, and $\delta_{\,\!\uparrow\!/\!\downarrow -}(t)=\omega_{\,\!\uparrow\!/\!\downarrow}(t)-\omega_{-}$. The eigenstates corresponding to the diagonal elements are
\begin{eqnarray}
|\Psi_1^{\!\uparrow\!/\!\downarrow}(t)\rangle &=& \frac{1}{\sqrt{2}}\left[\sqrt{1-\delta_{\,\!\uparrow\!/\!\downarrow -}(t)/[2\Omega_{\,\!\uparrow\!/\!\downarrow -}(t)]}|\!\!\uparrow\!/\!\downarrow \rangle- \sqrt{1+\delta_{\,\!\uparrow\!/\!\downarrow -}(t)/[2\Omega_{\,\!\uparrow\!/\!\downarrow -}(t)]}|-\rangle \right],\\
|\Psi_2(t)\rangle &=& \frac{1}{\sqrt{2}}\left[\sqrt{1+\delta_{\downarrow -}(t)/[2\Omega_{\downarrow -}(t)]}|\!\downarrow \rangle + \sqrt{1-\delta_{\downarrow -}(t)/[2\Omega_{\downarrow -}(t)]}|-\rangle\right],\\
|\Psi_3(t)\rangle &=& \frac{1}{\sqrt{2}}\left[\sqrt{1+\delta_{\uparrow -}(t)/[2\Omega_{\uparrow -}(t)]}|\!\uparrow \rangle + \sqrt{1-\delta_{\uparrow -}(t)/[2\Omega_{\uparrow -}(t)]}|-\rangle \right].
\end{eqnarray}
We note that $\omega_1^{\downarrow}(t)\approx \omega_1^{\uparrow}(t)$ throughout our whole range of the parameter $\omega_{\rm L}(t)$. Furthermore, $\omega_{2,3}(t)$ and the corresponding eigenstates $|\Psi_{2,3}(t)\rangle$ change slightly when the subspace of the diagonalization is changed, but with our parameters, the effect is negligible and omitted in the following discussion. Note also that the states $|\Psi_2(t)\rangle$ and $|\Psi_3(t)\rangle$ are approximately orthogonal since $\langle \Psi_2(t)|\Psi_3(t)\rangle = \sqrt{1-\delta_{\downarrow -}(t)/[2\Omega_{\downarrow -}(t)]}\sqrt{1-\delta_{\uparrow -}(t)/[2\Omega_{\uparrow -}(t)]}\approx 0$ for the relevant values of $\omega_{\rm L}(t)$. The Jacobian iteration could be continued further but by comparing these results with the numerical solution of the eigenproblem, we observe that the couplings between the states above are already so small that the Hamiltonian is accurately diagonalized. We have used these states in the derivation of the transition rates in Equations~(\ref{eq:transrate10})--(\ref{eq:transrate30}).


\subsection*{Ground-state fidelity during sudden sweeps}

The ground-state sweep fidelity $P_{\rm S}=|\langle \Psi_0(\tau)|\Psi_0(t_4)\rangle|^2$ can be calculated in the sudden approximation by finding the corrections to the RWA ground state $|\Psi_0^{\rm RWA}(t)\rangle = |0,0,{\rm g}\rangle$ caused by the counter-rotating terms $\hat H_2(t) = \hbar g_{\rm LR}(t) (\hat a^{\dag}_{\rm L} \hat a^{\dag}_{\rm R} + \hat a_{\rm L} \hat a_{\rm R}) +i\hbar g_{\rm Rq}(\hat a_{\rm R} \hat \sigma_- - \hat a_{\rm R}^{\dag}\hat \sigma_+)$. Since these change the occupation number by two, we should expand our low-energy basis to cover the eigenstates of the occupation numbers $N=0,1,2,3$ which gives altogether 16 basis states. However, the perturbation divides the eigenspace to two uncoupled sets, one of which is formed by the even-occupation states and the other one by the odd-occupation states. In conclusion, we do not expect the ground-state sweep fidelity to depend on the matrix elements between the ground state and any eigenstate with odd parity ($N=1,3,5,\ldots$). 
Thus, the counter-rotating terms break the symmetry in the RWA Hamiltonian that leads to the conservation of the occupation number, and replace it with a weaker requirement of parity conservation. 

We analytically calculate the correction arising from the counter-rotating term $\hat H_2(t)$ only for the ground state, which is created in our low-energy subspace by the off-resonant coupling between $|0,0,{\rm g}\rangle$ and the $N=2$ eigenstates. 
Since we have $g_{\rm LR}(t)/[\omega_{\rm L}(t)+\omega_{\rm R}]\ll 1$ and $g_{\rm Rq}/(\omega_{\rm q}+\omega_{\rm R})\ll 1$ during the whole protocol, we can treat the effects arising from $\hat H_2(t)$ in the second-order non-degenerate perturbation theory.
Since $H_2(t) |0,0,{\rm g}\rangle = g_{\rm LR}(t)|1,1,{\rm g}\rangle - ig_{\rm Rq}|0,1,{\rm e}\rangle$, we obtain
\begin{eqnarray}
E_0(t) &=& -\frac{g_{\rm LR}^2(t)}{\omega_{\rm L}(t)+\omega_{\rm R}}-\frac{g_{\rm Rq}^2}{\omega_{\rm q}+\omega_{\rm R}},\\
|\Psi_0(t)\rangle &=& A(t)\left(|0,0,{\rm g}\rangle-\frac{g_{\rm LR}(t)}{\omega_{\rm L}(t)+\omega_{\rm R}}|1,1,{\rm g}\rangle + i\frac{g_{\rm Rq}}{\omega_{\rm q}+\omega_{\rm R}}|0,1,{\rm e}\rangle\right),
\end{eqnarray}
where the zero of energy is set by the RWA ground-state energy and in the ground state we have neglected the second-order terms which are outside our low-energy subspace. The deviation from zero energy is caused by the Bloch--Siegert-type~\cite{Bloch40} non-resonant corrections to the RWA result. The normalization of the state is given by
\begin{eqnarray}
A(t) &=& \frac{[\omega_{\rm L}(t)+\omega_{\rm R}](\omega_{\rm q}+\omega_{\rm R})}{\sqrt{[\omega_{\rm L}(t)+\omega_{\rm R}]^2(\omega_{\rm q}+\omega_{\rm R})^2+g_{\rm LR}^2(t)[\omega_{\rm q}+\omega_{\rm R}]^2+g_{\rm Rq}^2[\omega_{\rm L}(t)+\omega_{\rm R}]^2}}\\
&\approx & 1-\frac{g_{\rm LR}^2(t)}{2[\omega_{\rm L}(t)+\omega_{\rm R}]^2}-\frac{g_{\rm Rq}^2}{2[\omega_{\rm q}+\omega_{\rm R}]^2},
\end{eqnarray}
where the latter equality is written to second order in the small parameters $g_{\rm LR}(t)/[\omega_{\rm L}(t)+\omega_{\rm R}]$ and $g_{\rm Rq}/(\omega_{\rm q}+\omega_{\rm R})$. Thus, the ground state can be approximately expressed as
\begin{equation}
|\Psi_0(t)\rangle \approx \left(1-\frac{g_{\rm LR}^2(t)}{2[\omega_{\rm L}(t)+\omega_{\rm R}]^2}-\frac{g_{\rm Rq}^2}{2[\omega_{\rm q}+\omega_{\rm R}]^2}\right)|0,0,{\rm g}\rangle-\frac{g_{\rm LR}(t)}{\omega_{\rm L}(t)+\omega_{\rm R}}|1,1,{\rm g}\rangle + i\frac{g_{\rm Rq}}{\omega_{\rm q}+\omega_{\rm R}}|0,1,{\rm e}\rangle.
\end{equation}
As a consequence, the sweep fidelity can be expressed as 
\begin{eqnarray}
P_{\rm S} 
& \approx & 
1- \left[\frac{g_{\rm LR}(\tau)}{\omega_{\rm L}(\tau)+\omega_{\rm R}}-\frac{g_{\rm LR}(t_4)}{\omega_{\rm q}+\omega_{\rm R}}\right]^2,\label{eq:ansurvpop}
\end{eqnarray}
where the sweep is assumed to start from $\omega_{\rm L}(t_4)=\omega_{-}\approx \omega_{\rm q}$. Thus, the maximum deviation from the perfect survival of the ground state is given by the square of the difference between the perturbation parameters.

\subsection*{Parameter optimization and the numerical method}

In this section, we optimize the parameters for our protocol. According to our analytic model, the decay time $\tau_{10}$ determines the transition rate as
\begin{equation}\label{eq:G0vstau10}
\Gamma_0=4\textrm{ln}(10)/\tau_{10}. 
\end{equation}
Furthermore, we require that in the off mode the 
transition rates are much smaller than the intrinsic rates $\kappa_{\rm R}^{\rm I}$ and $\kappa_{\rm q}^{\rm I}$. Thus, we obtain upper bounds for the transition rates in Equations~(\ref{eq:transrate10})--(\ref{eq:transrate30}) by writing them in the limit where $\omega_{\rm L}(0)\gg \omega_{\rm R},\omega_{\rm q}$ as~\cite{Jones13b}
\begin{eqnarray}
\Gamma_{10}^{\uparrow}(0) &\approx& \Gamma_0 \frac{(g_{\rm LR}^0)^2g_{\rm Rq}^2}{\delta_{\rm Lq}^2(0)\delta_{\rm Rq}^2}\left[\frac{(\delta_{\rm Lq}(0)+\omega_{\rm q})^2\omega_{\rm q}}{\omega_{\rm R}^3}\right] 
\ll \kappa_{\rm q}^{\rm I},\label{eq:ineqQ}\\
\Gamma_{20}(0) &\approx& \Gamma_0 \frac{(g_{\rm LR}^0)^2}{(\delta_{\rm Lq}(0)-\delta_{\rm Rq})^2}\frac{(\delta_{\rm Lq}(0)+\omega_{\rm q})^2}{\omega_{\rm R}^2} 
\ll \kappa_{\rm R}^{\rm I}, \label{eq:ineqR}\\
\Gamma_{30}(0) &\approx&\Gamma_0 \frac{\omega_{\rm L}^2(0)}{\omega_{\rm R}^2}
,
\end{eqnarray}
where we have employed the zero-temperature limit and $\delta_{\rm Lq}(0)$ determines the maximum sweep range in our protocol. If $\omega_{\rm L}(0)$ is large, the lowest three excited states can be approximated as $|\Psi_1(0)\rangle \approx |0,0,\rm e\rangle$, $|\Psi_2(0)\rangle \approx |0,1,\rm g\rangle$, and $|\Psi_3(0)\rangle \approx |1,0,\rm g\rangle$. We also note that $\Gamma_{20}(0)=\Gamma_{10}^{\uparrow}(0) \frac{\omega_{\rm R}\delta_{\rm Lq}^2(0)\delta_{\rm Rq}^2}{g_{\rm Rq}^2\omega_{\rm q}(\delta_{\rm Lq}(0)-\delta_{\rm Rq})^2}<\Gamma_{10}^{\uparrow}$, since the parameters of our protocol satisfy $\omega_{\rm q}<\omega_{\rm R}$ and $g_{\rm Rq}<\delta_{\rm Rq}$.

Another restriction for the transition rates comes from the fact that the master equation is valid only if
\begin{equation}
\Gamma_0\ll \min_{i\neq j}|\omega_i(t)-\omega_j(t)| = \left\{\begin{array}{ll}
\frac{g_{\rm LR}(t_3)g_{\rm Rq}}{\delta_{\rm Rq}}\approx \frac{g_{\rm LR}^0g_{\rm Rq}\sqrt{\omega_{\rm q}/\omega_{\rm R}}}{\delta_{\rm Rq}}, & \textrm{for } \omega_{\rm L}(t_3)=\omega_-\approx\omega_{\rm q};\\
g_{\rm LR}(t_1)\approx g_{\rm LR}^0, & \textrm{for } \omega_{\rm L}(t_1)=\omega_+\approx \omega_{\rm R},
\end{array}\right.\label{eq:Gammaineqs}
\end{equation}
where the latter approximations hold for our protocol since $g_{\rm LR}(t)=g_{\rm LR}^0\sqrt{\omega_{\rm L}(t)/\omega_{\rm R}}$. 
The first condition above determines the maximum $\Gamma_0$, since in our protocol $g_{\rm Rq}/\delta_{\rm Rq}<1$ and $\omega_{\rm q}/\omega_{\rm R} < 1$. In the following, we write 
\begin{equation}\label{eq:G0}
\Gamma_0 = \gamma\frac{g_{\rm LR}^0g_{\rm Rq}\sqrt{\omega_{\rm q}/\omega_{\rm R}}}{\delta_{\rm Rq}},
\end{equation}
and assume that $\gamma \ll 1$ so that inequalities~(\ref{eq:Gammaineqs}) hold.
We can optimize the decay time $\tau_{10}$ by defining the tolerances $\alpha_{\rm q} = \Gamma_{10}^{\uparrow}/\kappa_{\rm q}^{\rm I}$ and $\alpha_{\rm R} = \Gamma_{20}/\kappa_{\rm R}^{\rm I}$. The inequalities~(\ref{eq:ineqQ}) and~(\ref{eq:ineqR}) hold for $\alpha_{\rm q},\alpha_{\rm R}\ll 1$. We thus obtain from Equations~(\ref{eq:ineqQ}) and~(\ref{eq:ineqR}) that
\begin{equation}\label{eq:gLRineq}
g_{\rm LR}^0\leq \sqrt[3]{\frac{\alpha_{\rm q}\kappa_{\rm q}^{\rm I}\delta_{\rm Lq}^2(0)\delta_{\rm Rq}^3\omega_{\rm R}^3}{\gamma g_{\rm Rq}^3[\delta_{\rm Lq}(0)+\omega_{\rm q}]^2\omega_{\rm q}\sqrt{\omega_{\rm q}/\omega_{\rm R}}}} \times \min\left[1,
\sqrt[3]{\frac{\alpha_{\rm R}\kappa_{\rm R}^{\rm I}[\delta_{\rm Lq}(0)-\delta_{\rm Rq}]^2g_{\rm Rq}^2\omega_{\rm q}}{\alpha_{\rm q}\kappa_{\rm q}^{\rm I}\delta_{\rm Lq}^2(0)\delta_{\rm Rq}^2\omega_{\rm R}}}\right].
\end{equation}
We wish to minimize our decay time, which means that we need to maximize $g_{\rm LR}^0$. The largest allowed value for $g_{\rm LR}^0$ is given by the prefactor in the inequality above. This can be achieved when the qubit--right-resonator coupling is chosen as
\begin{equation}
g_{\rm Rq}\geq \frac{\delta_{\rm Lq}(0)\delta_{\rm Rq}}{|\delta_{\rm Lq}(0)-\delta_{\rm Rq}|}\sqrt{\frac{\alpha_{\rm q}\kappa_{\rm q}^{\rm I}\omega_{\rm R}}{\alpha_{\rm R}\kappa_{\rm R}^{\rm I}\omega_{\rm q}}}.
\end{equation}
In this case, an upper bound for the rate coefficient follows from Equations~(\ref{eq:G0}) and~(\ref{eq:gLRineq}), and can be written as
\begin{equation}
\Gamma_0 \leq \omega_{\rm R}\sqrt[3]{\frac{\gamma^2\alpha_{\rm q}\kappa_{\rm q}^{\rm I} \delta_{\rm Lq}^2(0)}{(\delta_{\rm Lq}(0)+\omega_{\rm q})^2\omega_{\rm R}}} \ll \sqrt[3]{\kappa_{\rm q}^{\rm I}\omega_{\rm R}^2},
\end{equation}
which is independent of $g_{\rm Rq}$. In the latter inequality above, we have assumed that $\omega_{\rm L}(0) \gg \omega_{\rm q}$. For typical experimental parameter values $\delta_{\rm Lq}(0)/(2\pi)=2$ GHz, $\omega_{\rm R}/(2\pi) = 10$ GHz, $\omega_{\rm q}/(2\pi) = 9.5$ GHz, $\delta_{\rm Rq}/(2\pi) = 0.5$ GHz, $\kappa_{\rm q}^{\rm I}=10^4$ s$^{-1}$, $\kappa_{\rm R}^{\rm I}=10^6$ s$^{-1}$, and for $\alpha_{\rm R}=\alpha_{\rm q}=0.1$, we obtain
\begin{eqnarray}
g_{\rm Rq}/(2\pi) &\geq & 68 \textrm{ MHz}; \\
g_{\rm LR}^0/(2\pi) &\leq & 74 \textrm{ MHz}; \\
\Gamma_0 &\leq & 31\times 10^6 \textrm{ s}^{-1}.
\end{eqnarray}
Above, we have set $\gamma=0.5$ and confirmed with a classical calculation of the transition rate $\Gamma_{20}(t_1)$ that in the case of $g_{\rm Rq}=0$, the secular approximation holds up to a relative error of $4\%$. 
The lower bound for the decay time is obtained from Equation~(\ref{eq:G0vstau10}) to be
\begin{equation}
\tau_{10}\geq 300 \textrm{ ns}.
\end{equation}

In our numerical simulations, we use the above-discussed values for the relevant parameters: $g_{\rm Rq}/(2\pi) = 68$ MHz, $g_{\rm LR}/(2\pi) = 74 $ MHz, $\Gamma_0 = 31 \times 10^6$ s$^{-1}$. These and the parameter values above are obtained, e.g., with circuit component values listed in Table~\ref{tab:params}.

\begin{table}[h]
\caption{Parameters used in the numerical simulations.}
\begin{tabular}{||*{6}{c|}|}
$C_{\rm R}=C_{\rm L}$ & $L_{\rm R}$ & $C_{\rm c}$ & $C_{\rm E}$ & $R$ & $T$ \\ \hline
1.0 pF & 250 pH & 15.0 fF & 4.0 fF & 500 $\Omega$ & 10 mK\\
\end{tabular}
\label{tab:params}
\end{table}

The numerically computed transition rates $\Gamma_{n0}(t)$ and eigenenergies $\omega_n(t)$ are obtained by solving the eigenproblem for $\hat{H}_{\rm S}(t)$ in the eigenbasis of the unperturbed Hamiltonian $\hat{H}_0(t)$, where we have included the eigenstates $|n_{\rm L\!\,/\,\!R}\rangle$ of the left and right-resonators up to $n_{\rm L\!\,/\,\!R}=5$, and solved the master equation in the static basis consisting of five energetically lowest states. We have confirmed that the relative error in the final ground-state occupation caused by the truncation of the basis is of the order of $10^{-7}$ or smaller.

\section*{Acknowledgements}

This work was supported by the Academy of Finland through its Centers of Excellence Programme under project numbers 251748 and 284621 and by the European Research Council under Starting Independent Researcher Grant number 278117 (SINGLEOUT) and Consolidator Grant number 681311 (QUESS).



\section*{Competing interests}


The authors declare no conflict of interest.


\section*{Contributions}

J.T. carried out the analytic and numerical calculations based on the initial idea of M.M., and was mainly responsible for writing the manuscript. M.P. made the classical calculation of the transition rates. T.A.-N. and M.M. provided feedback and supervised the project. All authors discussed the results and commented on the manuscript.





\bibliographystyle{naturemag}
\bibliography{QubitInit}

\end{document}